\documentstyle[12pt]{article}
\newlength{\dinwidth}
\newlength{\dinmargin}
\makeatletter
\def\fmslash{\@ifnextchar[{\fmsl@sh}{\fmsl@sh[0mu]}}
\def\fmsl@sh[#1]#2{%
  \mathchoice
    {\@fmsl@sh\displaystyle{#1}{#2}}%
    {\@fmsl@sh\textstyle{#1}{#2}}%
    {\@fmsl@sh\scriptstyle{#1}{#2}}%
    {\@fmsl@sh\scriptscriptstyle{#1}{#2}}}
\def\@fmsl@sh#1#2#3{\m@th\ooalign{$\hfil#1\mkern#2/\hfil$\crcr$#1#3$}}
\makeatother
\global\arraycolsep=2pt 
\input{epsf}
\begin{document}
\thispagestyle{empty}
\begin{titlepage}
\hskip -3cm
\begin{flushright}
{\bf TTP 96 -- 23}
\end{flushright}


\begin{center}
{\Large\bf Effective Theory for Heavy Quarks}
\end{center}

\vspace{0.8cm}

\begin{center}
{\sc Thomas Mannel}  \vspace*{2mm} \\
{\sl Institut f\"{u}r Theoretische Teilchenphysik,
     University of Karlsruhe \vspace*{1mm} \\
     Kaiserstra\ss e 12, D -- 76128 Karlsruhe, Germany.}
\vspace*{2cm} \\
{\it Lectures given at the \\
35. Internationale Universit\"atswochen f\"ur Kern- und Teilchenphysik, \\
March 2--9, 1996, Schladming, Austria}
\end{center}
\vfill
\begin{abstract}
\noindent
In this series of lectures the basic ideas of the $1/m_Q$ 
expansion in QCD ($m_Q$ is the mass of a heavy quark) are 
outlined. Applications to exclusive and inclusive decays 
are given. 
\end{abstract}
\tableofcontents
\end{titlepage}
\newpage
\section{Introduction}
With the precise formulation of the $1/m_Q$ expansion in QCD
using effective field theory and operator product expansion 
heavy quark physics has been based on model independent ground, 
which allows to reduce the uncertainties due to the QCD bound 
state problem drastically. From the point of view of weak 
interactions the main interest in processes with heavy quarks 
lies in the exploration and the test of the CKM sector of the 
standard model (SM) describing the masses and the mixing of quarks. 
From the experimental side 
the running experiments as well as the ones planned in the near 
future constitute a large effort to explore this part of the SM, 
which is not yet tested with an accuracy comparable with the 
one of the coupling of the $Z_0$ to the fermions.   

The main theoretical progress in the description of systems involving
a single heavy quark is based on the infinite mass limit of QCD 
\cite{shifman}, in which two additional symmetries appear that 
are not present in full QCD \cite{isgur/wise}. This limit may 
be regarded as the leading term of a $1/m_Q$ expansion and a 
systematic approximation to full QCD may be constructed using the 
methods of effective field theory, the 
so-called Heavy Quark Effective Theory (HQET) \cite{HQET}.  
The heavy mass limit and applications of HQET 
have been extensively studied, and 
the development of the field is documented in more or less extensive 
reviews \cite{reviews}. 

The corrections to the heavy mass limit are characterized by two small 
parameters, namely the strong coupling constant,
taken at the scale of the heavy quark $\alpha_s (m_Q)$ and the 
ratio $\bar\Lambda / m_Q$ of the scale of the light degrees of 
freedom $\bar\Lambda$ and the heavy quark mass. While the first 
kind of corrections may be calculated perturbatively in terms of 
Feynman diagrams, the second type needs additional non-perturbative input
encoded in the matrix elements of higher dimensional operators. 

However, the non-perturbative input is to be taken at a small scale, 
where the symmetries of HQET may be applied and hence the 
non-perturbative physics is constrained by these symmetries. 
In particular, for heavy to heavy decays these additional symmetries
restrict the number of independent form factors to only a single one, 
of which the absolute normalization at a specific kinematic point is 
fixed by the symmetries.

By combining the method of the $1/m_Q$ expansion with the 
short distance expansion one may obtain a heavy mass expansion 
also for inclusive decay rates
\cite{Manifesto}-\cite{BigiPRL},\cite{CCGincsl}-\cite{FLS93}.
The heavy quark mass 
sets a scale that is large compared to $\Lambda_{QCD}$, 
and one may use a similar
setup as in deep inelastic scattering for the description of 
inclusive decays. In this
way one may not only study total rates, but also differential distributions
such as the lepton energy spectra in inclusive semileptonic decays. 

This series of three  lectures tries to summarize the basic ideas of the 
heavy mass expansion. In the first lecture the formulation of HQET 
as an effective field theory is described and the additional symmetries 
of heavy quarks are introduced. The second and third lecture 
deal with applications to exclusive and inclusive decays.

\section{Effective Field Theory}
Effective field theories \cite{effth} have become a widely used tool in 
modern elementary particle physics. An effective theory treatment 
is convenient if the problem under consideration involves very 
disparate mass scales such that the physics that is to be described 
happens at much lower energies than the scale set by some heavy 
particles in the theory. In such a case it is useful to switch to 
an effective theory in which the heavy degrees of freedom do not 
appear explicitly; they only reappear in the
effective theory as higher dimensional operators, which 
are multiplied by coupling constants with negative mass dimension. 
The scale of the coupling constants is set by the large mass  
and thus these contributions are small, if the scale of the physics
described with the help of the effective theory is small compared to 
this large mass. 

An effective theory is always valid only in a limited region of scales, 
a natural cut-off is given by the mass of the particle which has been 
removed by switching from the full to the effective theory. As 
mentioned above an effective theory involves interactions which would 
lead to a non-renormalizable theory, if one would consider the theory 
to all orders in these higher dimensional operators. However, working 
to a definite order of the expansion in inverse powers of the large scale 
one does not face any problem concerning renormalization. Starting from   
the renormalizable dimension-4 piece of the effective theory 
Lagrangian we may use its renormalization group properties to study 
the cut-off dependence of the effective theory, which is determined by 
the short distance properties of the effective theory. 

Applying effective theory methods corresponds to an expansion of the 
Greens functions of the full theory in inverse powers of the large mass 
scale; such an expansion is only possible up to logarithmic dependences
on this large scale. These logarithms may be accessed
using a properly constructed effective theory, where these logarithms 
correspond to the renormalization group logarithms of the cut off. In this 
way one may even achieve a resummation of the logarithmic terms using 
renormalization group methods in the effective theory.  

In the case at hand the large scale is the mass $m_Q$ of the 
heavy quark, and the leading term of this expansion is the static limit. 
This effective theory (Heavy Quark Effective Theory, HQET) is a 
a powerful tool, allowing for numerous 
purely QCD based calculations. Renormalization in this effective
theory implies a factorization theorem for
the Greens function of full QCD, which means that to any
order in the $1/m_Q$ expansion one may factorize the short distance physics
from the long distance effects. Explicitly this means for a Greens function 
$G_{full}$ calculated in full QCD that one may rewrite it as
\begin{equation} \label{fact}
G_{full} (p_1 \cdots p_n, m_Q, \mu_0 = m_Q) = \sum_j \left(\frac{1}{m_Q}\right)^j
Z^{(j)} (m_Q,\mu) G_{eff}^{(j)} (p_1 \cdots p_n, \mu)  .
\end{equation}
where $\mu_0$ is the renormalization point of the full QCD function.  
Here the constants $Z^{(j)}(m_Q,\mu)$ depend on the factorization scale $\mu$
and on logarithms of $m_Q$; it 
contains all the short distance effects, which may be calculated 
perturbatively. Using the renormalization group of the effective theory, 
one may perform a systematic resummation of logarithmic dependences on the
heavy quark mass. 

The Greens functions $G^{(j)}_{eff}$ are calculated in HQET and do not depend on 
the heavy mass any more; 
they contain the long distance effects which are not calcualable via perturbation 
theory. However, as we shall see below, they are constrained by heavy quark 
symmetries. 
Consequently one gains complete control over the mass dependence by switching 
from full QCD to HQET, the effective theory obtained in the heavy mass limit.   

In this section we shall discuss the formulation of this effective theory. 
First  we shall review and compare some of the different possibilities to 
formulate the infinite mass limit. In this limit new symmetries appear which 
are the key to various model independent statements concerning weak 
decay matrix elements; these symmetries are reviewed in paragraph 3.1.
Finally, we shall consider the systematic approach to the calculation of 
corrections to the infinite mass limit. 
  
\subsection{Infinite Quark Mass Limit as an Effective Field Theory}
This issue has been discussed repeatedly and various formulations
of the infinite mass limit are available. Of course, as far as physical 
quantities are concerned, all approaches yield the same result. However, 
for some special applications one approach may be more convenient than 
another. 

The equivalence of all different approaches is ensured by a theorem 
well established in the field  of effective theories. It has been 
shown that in an effective theory involving a field $\phi$ one may 
perform redefinitions of the fields , such that 
\begin{equation}
\phi \to P (\phi, \partial \phi) 
\end{equation}
where $P$ is an arbitrary polynomial function \cite{CCWZ,KO94}.  

Such a redefinition will not change the $S$-matrix, although the Lagrangian
(and Greens function) expressed in the redefined fields may look completely 
different. In this sense the different formulations of HQET are equivalent.   

In the following we shall consider two formulations of HQET. The first one 
nicely exhibits the fact that HQET is an effective theory in the sense that 
one integrates out a heavy degree of freedom and performs an expansion of 
the remaining action functional in the large mass scale, which for the 
case at hand is the heavy quark mass. The process of integrating out the heavy 
degree of freedom may be performed explicitly as a Gaussian functional 
integral, and one may construct a formulation of HQET be expanding the 
result in powers of $1/m_Q$. In this way one obtains a $1/m_Q$ expansion 
of both the heavy quark field as well as for the Lagrangian.

The second formulation is based on the standard way of separating ``upper''
and ``lower'' components of the spinor fields by performing a sequence of 
Foldy Wouthuysen transformations. These transformations lead to an expansion 
in $1/m_Q$ for the heavy quark field and for the Lagrangian, which is 
different for each of these quantities from the other formulation. However, 
if one calculates a physical quantity, both approaches will yield the same 
answer, because the $1/m_Q$ expansion has to be unique for an observable. 

\subsubsection*{Integrating out heavy degrees of freedom}
One may obtain a formulation of the heavy mass limit by integrating out heavy 
degrees of freedom from the functional integral of QCD Greens functions 
\cite{MRRderivation}. This 
integration may in fact be done explicitly, since for the case at hand it amounts 
to a Gaussian functional integration. Formulating the heavy mass limit in this 
way clearly exhibits that it corresponds to an effective theory in the usual sense. 
We start from the generating functional of the QCD Greens functions
\begin{equation} \label{gf}
Z(\eta,\bar{\eta},\lambda) = \int [d Q][d \bar{Q}]
                                  [d \phi_\lambda ]
\exp \biggl\{ i S + i S_\lambda
     + i \int d^4 x \, (\bar{\eta} Q + \bar{Q} \eta
                        + \phi_\lambda \lambda )
     \biggr\},
\end{equation}
where $\phi_\lambda = q,\, A_\mu^a$ denotes the light degrees of freedom
(light quarks $q$ and gluons $A_\mu$) with the action $S_\lambda$, while 
$S$ denotes the piece of the action for the heavy quark $Q$ including its coupling
to the gluons 
\begin{equation} \label{act}
S = \int d^4 x \, \bar{Q} (i \fmslash{D} - m_Q ) Q.
\end{equation}
where 
\begin{equation}
D_\mu = \partial_\mu + i g A_\mu, \quad
\end{equation}
is the covariant derivative of QCD. We have introduced source terms 
$\eta$ for the heavy quark and $\lambda$ for the light degrees of freedom. 

We shall consider hadrons containing a single heavy quark, and we assume 
that this heavy hadron moves with a velocity $v$
\begin{equation}
v = \frac{p_{\mbox{hadron}}}{m_{\mbox{hadron}}}, \quad v^2 = 1, \quad v_0 > 0
\end{equation}
This velocity vector may be used to 
split the heavy quark field $Q$  into an ``upper'' component $\phi$ and a
``lower'' one $\chi$
\begin{eqnarray} \label{splitu}
\phi_v = \frac{1}{2} (1+\fmslash{v}) Q, && \fmslash{v} \phi_v = \phi, \\
\chi_v = \frac{1}{2} (1-\fmslash{v}) Q, && \fmslash{v} \chi_v = -\chi , 
\label{splitl}
\end{eqnarray}
and to define a decomposition of the covariant derivative into a
``longitudinal'' and a ``transverse'' ($\perp$)
part
\begin{equation} \label{splitd}
D_\mu = v_\mu (v \cdot D) + D_\mu^{\perp} \, , \quad
D^{\perp}_\mu  = (g_{\mu \nu} - v_\mu v_\nu ) D^\nu, \quad
\left\{ \fmslash{D}^{\perp} \, , \, \fmslash{v} \right\} = 0.
\end{equation}
Using (\ref{splitu}-\ref{splitd}) the action
(\ref{act}) of the heavy quark field takes the
form
\begin{equation} \label{actul}
S = \int d^4 x \, \biggl[
    \bar{\phi} \{ i (v \cdot D) - m_Q \} \phi -
    \bar{\chi} \{ i (v \cdot D) + m_Q \} \chi 
+ \bar{\phi} i \fmslash{D}^{\perp} \chi
    + \bar{\chi} i \fmslash{D}^{\perp} \phi \biggr].
\end{equation}
The heavy quark in the meson is very close to being on shell, and thus
the space time dependence of the heavy quark field is mainly that
of a free particle moving with velocity $v$. This suggests a
reparametrization of the fields by removing
the space time dependence of a solution of the free Dirac equation. 
We shall choose the ``particle-type'' parametrization corresponding to 
the ``positive energy solution'' of the Dirac equation 
\begin{equation} \label{phaseredef}
\phi_v = e^{-im_Q (v \cdot x)} h_v \, , \qquad
\chi_v = e^{-im_Q (v \cdot x)} H_v,
\end{equation}
such that the space time dependence of the remaining fields $h_v$ and 
$H_v$ is determined by the residual momentum $k = p - m_Q v$, which is 
due to binding effects of the heavy quark inside the heavy hadron, and 
which is a ``small'' quantity of order $\Lambda_{QCD}$.  

Expressed in these fields the action of the heavy quark becomes 
\begin{equation} \label{acth}
S = \int d^4 x \, \left[
    \bar{h}_v i (v \cdot D) h_v -
    \bar{H}_v \{ i (v \cdot D) + 2m_Q \} H_v 
    + 
      \bar{h}_v i \fmslash{D}^{\perp} H_v
    + \bar{H}_v i \fmslash{D}^{\perp} h_v \right],
\end{equation}

The term containing the sources is also rewritten in terms of the
fields $h_v$ and $H_v$
\begin{equation}
 \int d^4 x \, (\bar{\eta} \psi + \bar{\psi} \eta)
 =  \int d^4 x \, (\bar{\rho}_v h_v + \bar{h}_v \rho_v
             + \bar{R}_v H_v + \bar{H}_v R_v ),
\end{equation}
where $\rho_v$ and $R_v$ are now source terms for the upper component 
field $h_v$ and the lower componenent part $H_v$ respectively. 

In terms of the new variables the generating functional reads
\begin{eqnarray} \label{Genfh}
&& Z ( \rho_v ,  \bar{\rho}_v , R_v , \bar{R}_v, \lambda )
=  \int [d h_v][d \bar{h}_v][d H_v][d \bar{H}_v]
        [d \phi_\lambda ] \\
&&   \exp \left\{ i S +  S_\lambda+
    i \int d^4 x \, (\bar{\rho}_v h_v + \bar{h}_v \rho_v
             + \bar{R}_v H_v + \bar{H}_v R_v
             + \phi_\lambda \lambda )
          \right\},
\nonumber
\end{eqnarray}
where the action $S$ for the heavy quark is given in eq.(\ref{acth}).

From (\ref{acth}) it is obvious that the 
heavy degree of freedom is the lower component field $H_v$, since it has
a mass term $2m_Q$, while the upper component field $h_v$ is a massless 
field describing the static heavy quark. In the heavy mass limit only 
the Greens functions involving the field $h_v$ have to be calculated, 
and hence we integrate over $H_v$ in the functional integral 
(\ref{Genfh}) with the sources of the lower component field $R_v$ and 
$\bar{R}_v$ set to zero. This can be done explicitly, since it is a Gaussian 
integration 
\begin{equation}
Z ( \rho_v ,  \bar{\rho}_v , \lambda )
=
     \int [d h_v][d \bar{h}_v]
          [d \lambda ] \Delta 
   \exp \left\{ i S + S_\lambda +
     i \int d^4 x \, (\bar{\rho}_v^+ h_v^+ + \bar{h}_v^+ \rho_v^+
                      + \phi_\lambda \lambda )
       \right\},
\end{equation}
where now the action functional for the heavy quark becomes a non-local object 
\begin{equation}  \label{actnl}
S = \int d^4 x \,  \left[
    \bar{h}_v^+ i (v \cdot D) h_v^+
    - \bar{h}_v^+  \fmslash{D}^{\perp}
    \left( \frac{1}{ i (v \cdot D) + 2m_Q - i \epsilon} \right)
    \fmslash{D}^{\perp} h_v^+ \right].
\end{equation}
This Gaussian integration corresponds to the replacement 
\begin{equation}
H_v = \left( \frac{1}{2m_Q + ivD} \right) i \fmslash{D}_\perp h_v
\end{equation}
for the lower compponent field.
Furthermore, the Gaussian integration yields a determinant 
$\Delta$. In the full theory one may also perform this Gaussian 
integration, and the determinant obtained contains all the closed 
loops of heavy quarks. After renormalization of the full theory their 
contribution starts at order $1/m^2$ with an Uehling potential like term.
In the effective theory one may take the determinant $\Delta$ to be a 
constant, if the terms of order $1/m_Q^2$ and higher comming from the 
closed heavy quark loops are included by matching to the full theory. 
Since we shall discuss only the leading term of the $1/m_Q$ expansion 
in this section, we may drop the determinant in what follows.    

The non-locality of the action functional is connected to the large scale 
set by the heavy quark mass, and the non-local terms may be expanded in 
terms of an infinite series of local operators, which come with increasing
powers of $1/m_Q$. In the context of a field theory this corresponds 
to a short distance expansion and hence these operators have to be 
renormalized. The tree level relations may be read off from the geometric 
series expansion of the non-local term in (\ref{actnl}). In this way we 
obtain the expansion of the field and the Lagrangian 
\begin{eqnarray} \label{HarvField}
Q(x) &=& e^{-im_Qvx} \left[ 1 + 
         \left( \frac{1}{2m + ivD} \right) i \fmslash{D}_\perp \right] h_v \\ 
     &=& e^{-im_Qvx} \left[ 1 + \frac{1}{2m_Q} \fmslash{D}_\perp +  \nonumber 
     \left(\frac{1}{2m_Q}\right)^2 (-ivD) \fmslash{D}_\perp + \cdots \right] h_v \\  
\label{HarvLag}
{\cal L} &=& \bar{h}_v (iv D) h_v + 
\bar{h}_v i \fmslash{D}_\perp \left( \frac{1}{2m + ivD} \right) i \fmslash{D}_\perp h_v \\
         &=& \bar{h}_v (iv D) h_v + \frac{1}{2m} \bar{h}_v (i\fmslash{D}_\perp)^2 i h_v
          +  \left( \frac{1}{2m}\right) \bar{h}_v (i\fmslash{D}_\perp) (-ivD) 
         \nonumber                               (i\fmslash{D}_\perp)  h_v + \cdots
\end{eqnarray}

A Greens function with an operator insertion is treated in a similar way; the 
heavy quark fields entering the inserted operator are dealt with in the 
same way.
The net effect of this is that the heavy quark fields in the operator 
insertion are replaced by the expansion (\ref{HarvField}).

\subsubsection*{Foldy Wouthuysen Transformation}
A second way of formulating the heavy mass limit proceed along the well 
known steps performed in deriving the non-relativistic limit of the Dirac 
equation \cite{KoernerFW}. The reasoning used here is motivated by 
quantum mechanics; as 
a first step one rewrites the equation of motion in a Hamiltonian form 
\begin{equation} \label{schroe}
iv \partial Q = H Q, \quad 
H = \fmslash{v} (\fmslash{D}_\perp + m_Q + g vA ) 
\end{equation}   
where $A$ is the gluon field. Note that in the rest frame we have 
$v = (1,0,0,0)$, and thus (\ref{schroe}) takes the usual form, 
since $v \partial = \partial_0$ and $vA = A_0$.    

In general the Hamiltonian couples the upper and the lower component 
of the heavy quark field $Q$, projected out by 
$P_\pm = (1 \pm \fmslash{v})$. The Foldy Wouthuysen transformation
is a transformation of the form  
\begin{equation} \label{FW}
Q \to Q' = \exp(iF) Q \quad 
H \to H' = \exp(iF) \left[H - i v \partial \right] \exp(-iF)
\end{equation}
where $F$ is a hermitian matrix in the space of the Dirac spinors, 
such that $\exp(iF)$ is unitary in that space. This requirement is 
motivated by quantum mechanics where the spinor is interpreted as a 
wave function, but this interpretation becomes meaningless once we 
switch to a field theory.  

The transformation (\ref{FW}) is designed such that the resulting 
Hamiltonian $H'$ does not couple upper and lower components of the 
field $Q$ any more. The generator $F$ of 
this transformation may be expanded in powers of $1/m_Q$, from which 
one may construct a $1/m_Q$ expansion of the Hamiltonian and the 
transformed fields. Removing the mass term of the Hamiltonian by a phase 
redefinition as in (\ref{HarvField}), one obtains for the fields and 
the Lagrangian
\begin{eqnarray} \label{FWfield}
&& Q(x) =  e^{-im_Qv\cdot x}\left[ 1 +\frac{1}{2m_Q}(i\fmslash{D}_{\perp}) +
\frac{1}{4m_Q^2}\left( v\cdot D \fmslash{D}_{\perp} - \frac{1}{2}
\fmslash{D}_{\perp}^2 \right) + \cdots \right] h_v (x)\,,
\\ \label{FWlag}
&& {\cal L} = \bar h_v\left[ iv \cdot D - \frac{1}{2m_Q}
\fmslash{D}_{\perp}^2
 + \frac{i}{4m_Q^2}\left(-\frac{1}{2}
\fmslash{D}_{\perp}^2 v \cdot D +
\fmslash{D}_{\perp} v\cdot D 
\fmslash{D}_{\perp} - \frac{1}{2} v\cdot D 
\fmslash{D}_{\perp}^2 \right) + \cdots \right] h_v\,.\nonumber\\
\end{eqnarray}
We note that the leading order term as well as the terms of order $1/m_Q$ 
are identical in the two approaches. Differences start to appear at order 
$1/m_Q^2$, which are terms involving a factor which would vanish 
by leading order equations of motion. In the Fouldy Wouthuysen formulation 
the Lagrangian does not contain such terms, while these terms appear in the 
approach of integrating out the lower components of the heavy quark field. 
However, subleading terms of a physical matrix element will consist of
local contributions originating from the expansion of the field $Q$ 
as well as of non-local pieces involving time-ordered products of the leading 
order currents with the subleading terms of the Lagrangian (see below).
If the time-ordered products are taken with a term that would vanish by a 
naive 
application of the leading-order equation of motion, these will lead to a 
contact terms, i.e.\ effectively to a local contribution. In this way 
the terms in the first approach rearrange in such a way that the final result 
for an observable quantity is the same in both cases. For some practical
applications the Foldy Wouthuysen approach has an advantage that all the 
time-ordered product with the Lagrangian are truly non-local contributions, 
in other words, none of the contributions  will lead to a contact term.        

\subsection{Corrections to the Heavy Mass Limit}
\subsubsection*{Tree Level Considerations}
Corrections to the infinite mass limit may be considered in a systematic way. 
Starting from the tree level expressions given in the last sections one may 
use the expansion of the Lagrangian and the fields as given in section 2.1. 
to construct the $1/m_Q$ expansion of full QCD matrix elements. In doing 
this it will not matter which representation of HQET (e.g. the one that is 
obtained from integrating out the heavy quark or the one constructed from 
the Foldy Wouthuysen transformation), since the matrix elements have to have a
unique $1/m_Q$ expansion. 

As an example we shall consider a matrix element of a current 
$\bar{q} \Gamma Q$ mediating
a transition between a heavy meson and some arbitrary state 
$| A \rangle$. The full QCD Lagrangian ${\cal L}$ and the fields $Q$ 
are expanded in terms of a power series in $1/m_Q$ in the way described in 
section 2.1, and the matrix element under consideration 
up to order $1/m_Q$ takes the form:
\begin{eqnarray} \label{exp}
&& \langle A | \bar{q} \Gamma Q 
              | M (v) \rangle =  
\langle A | \bar{q} \Gamma h_v 
              | H (v) \rangle  \\ \nonumber 
&& \qquad + \frac{1}{2m_Q} \langle A | \bar{q} \Gamma 
              P_- i \fmslash{D} h_v 
              | H (v) \rangle    
 -i \int d^4 x \langle A | T \{ L_1 (x) \bar{q} \Gamma h_v \}
              | H (v) \rangle   
   + {\cal O} (1/m^2)  
\end{eqnarray}
where $L_1$ are the first-order corrections to the Lagrangian
as given in (\ref{HarvLag}) or (\ref{FWlag}). Furthermore, $| M (v) \rangle $ is the 
state of the heavy meson in full QCD, including all its mass 
dependence, while $| H (v) \rangle$ is the corresponding state in 
the infinite mass limit. 

Expression (\ref{exp}) displays the generic structure of the 
higher-order corrections as they appear in any HQET calculation. 
There will be local contributions coming from the expansion of 
the full QCD field; these may be interpreted as the corrections to 
the currents. The non-local contributions, i.e.~the time-ordered
products, are the corresponding corrections to the states and thus
in the r.h.s.\ of (\ref{exp}) only the states of the infinite-mass 
limit appear.  If one switches to another representation 
of HQET, one reshuffles 
terms from the fields into the Lagrangian; in this way the Lagrangian 
picks up operators which are proportional to the equations of motion.
  
As an example for the kinds of matrix elements appearing in subleading 
orders of the $1/m_Q$ expansion 
we consider the mass of a heavy hadron. 
In the infinite mass limit this mass is given in terms
of the quark mass plus some ``binding energy'' 
$\bar{\Lambda}$. The corrections are of order $1/m_Q$
and are given by the matrix elements of the leading 
correction term of the Lagrangian. One obtains 
\cite{mass}
\begin{equation} \label{massrel}
m_H = m_Q \left(1 + \frac{\bar\Lambda}{m_Q} + \frac{1}{2 m_Q^2} 
      \left(\lambda_1 + d_H \lambda_2 \right)
+ {\cal O} (1/m_Q^3) \right)  
\end{equation}
where $d_H = 3$ for the $0^-$ and $d_H = -1$ for the $1^-$ meson. Up to 
this order no non-local terms appear; such terms show up the first time 
at order $1/m_Q^3$. 
The parameters $\bar\Lambda$, $\lambda_1$ and $\lambda_2$ correspond to 
matrix elements involving higher order terms that appear in the effective
theory Lagrangian 
\begin{eqnarray} 
\bar{\Lambda} &=& 
\frac{\langle 0 | q \stackrel{\longleftarrow}{ivD} \gamma_5 h_v | H (v) \rangle}
                       {\langle 0 | q \gamma_5 h_v | H (v) \rangle}
\label{Lambar} \\ 
\lambda_1 &=& \frac{\langle H (v) | \bar{h}_v  (iD)^2  h_v | H (v) \rangle}
                   {2 M_H} 
\label{lam1} \\
\lambda_2  &=& \frac{\langle H (v) | \bar{h}_v \sigma_{\mu \nu} iD^\mu iD^\nu h_v 
                                   | H (v) \rangle}
                    {2 M_H}
\label{lam2}
\end{eqnarray}
where the normalization of the states is chosen to be 
$\langle H (v) | \bar{h}_v h_v | H (v) \rangle = 2 M_H = 2 (m_Q + \bar\Lambda) $. 
These parameters may be interpreted as the binding energy of the heavy 
meson in 
the infinite mass limit ($\bar\Lambda$), the expectation value of 
the kinetic energy  of the heavy quark ($\lambda_1$) and its energy 
due to the chromomagnetic moment of the heavy quark ($\lambda_2$)
inside the heavy meson. The latter two parameters play an important role since 
they parametrize the non-perturbative input needed in the subleading order of the 
$1/m_Q$ expansion.   

The only parameter which is easy to access is
$\lambda_2$, since it is related to the mass splitting between $H(v)$ 
and $H^* (v , \epsilon)$. From the $B$-meson system we obtain
\begin{equation}
\lambda_2 (m_b) = \frac{1}{4} (M_{H^*} - M_H) = 0.12 \mbox{ GeV}^2; 
\end{equation}
from the charm system the same value is obtained.  This shows that indeed
the spin-symmetry partners are degenerate in the infinite mass limit and 
the splitting between them scales as $1/m_Q$. 

The other parameters appearing in (\ref{massrel}) are not simply 
related to the hadron spectrum.
Using the pole mass for $m_Q$ in (\ref{massrel}), QCD sum rules yield for
a value of $\bar\Lambda = 570 \pm 70$ MeV \cite{reviews}. More 
problematic is the parameter $\lambda_1$; from its definition one 
is led to assume
$\lambda_1 < 0$; a more restrictive inequality 
\begin{equation}
-\lambda_1 > 3 \lambda_2
\end{equation}
has been derived in a quantum mechanical framework in \cite{BiMotion}
and using heavy-flavour sum rules \cite{Bisumrule}.       
Furthermore, there exists also a sum rule estimate \cite{BBsumrule}
for this parameter:
\begin{equation}
\lambda_1  = - 0.52 \pm 0.12 \mbox{ GeV}^2 .
\end{equation}
This value is compatible with the bounds; however, it is unexpectedly large
since it corresponds to a rms-momentum of the heavy quark inside the meson 
of 
\begin{equation}
\sqrt{\langle \vec{p}^2 \rangle} \sim 720 \mbox{ MeV} 
\end{equation}
which is large compared to the naive guess of $-\lambda_1 \sim (\Lambda_{QCD})^2$ 
This is the reason why also smaller values of $\lambda_1$ have been used 
in the literature. 

Recently there has been an attempt \cite{GKLW96} to extract 
$\bar\Lambda$ and $\lambda_1$ from the shape of the lepton energy spectrum 
in inclusive semileptonic 
$B$ decays (see sections 4.3 and 4.4). The values obtained from this analysis are
$\bar\Lambda = 0.39 \pm 0.11$ GeV and $-\lambda_1 = 0.19 \pm 0.10$ GeV${}^2$, 
where the $\overline{\rm MS}$ definition of the mass has been used. 
The uncertainties
quoted are only the $1\sigma$ statistical ones; the systematical uncertainties 
of this approach are difficult to estimate.  

\subsubsection*{Beyond Tree Level}

Going beyond tree level will induce corrections of order $\alpha_s^n (m_Q)$, 
$n = 1, ...$. These may be calculated in terms of Feynman diagrams which 
may be evaluated using the Feynman rules of HQET. Only two Feynmal rules are 
modified compared to full QCD:

\begin{center}
\begin{tabular}{llll}
        & full QCD & & HQET \\
Propagator of the heavy quark
& $\displaystyle{\frac{i}{\fmslash{p} - m_Q + i\epsilon}}$ & $\longrightarrow$
& $\displaystyle{\frac{i}{vk + i\epsilon}}, \quad p = mv + k $ \\
Heavy quark gluon vertex
& $ ig \gamma_\mu T^a $ & $\longrightarrow$
& $ ig v_\mu T^a $ \\
& & & 
\end{tabular}
\end{center}

For the sake of clarity we shall stick to our example of a heavy light 
current considered above. To leading order in the $1/m_Q$ expansion one 
may evaluate the radiative corrections to such a matrix element using the 
above Feynman rules and finds a divergent result with a divergence related 
to the short distance behaviour. Since HQET is an 
effective theory, the machinery of effective theory guarantees the 
factorization of long distance effects from the short distance ones, 
which are related to the large mass $m_Q$.  
Neglecting $1/m_Q$ corrections, this factorization takes the form
\begin{equation} \label{expr}
\langle A | \bar{q} \Gamma Q 
              | M (v) \rangle = Z \left( \frac{m_Q}{\mu} \right)  
\langle A | \bar{q} \Gamma h_v 
              | H (v) \rangle |_\mu 
   + {\cal O} (1/m_Q)  
\end{equation}
From Feynman rule calculation one obtains the perturbative expansion 
of the renormalization constant $Z$ which generically looks like  
\begin{eqnarray} \label{zexp}
Z \left( \frac{m_Q}{\mu} \right) &=& a_{00} \\
&+& a_{11} \left( \alpha_s \ln \left(\frac{m_Q}{\mu} \right) \right)\,
 +  a_{10} \alpha_s \nonumber \\ 
&+& a_{22} \left( \alpha_s \ln \left(\frac{m_Q}{\mu} \right) \right)^2
 +  a_{21} \alpha_s
           \left( \alpha_s \ln \left(\frac{m_Q}{\mu} \right) \right)
 +  a_{20} \alpha_s^2 \nonumber \\
&+& a_{33} \left( \alpha_s \ln \left(\frac{m_Q}{\mu} \right) \right)^3
 +  a_{32} \alpha_s
           \left( \alpha_s \ln \left(\frac{m_Q}{\mu} \right) \right)^2
 +  a_{31} \alpha_s^2
           \left( \alpha_s \ln \left(\frac{m_Q}{\mu} \right) \right) 
\nonumber \\
&& \hphantom{--------} +  a_{30} \alpha_s^3 
 + \cdots \nonumber
\end{eqnarray}
where $\alpha_s = g^2/(4\pi)$.

This factorization theorem corresponds to the statement that the
ultraviolet divergencies in the effective theory have to match the
logarithmic mass dependences of full QCD. The factorization scale $\mu$ is
an arbitrary parameter, and the physical quantity
$\langle A | \bar{q} \Gamma Q | M (v) \rangle $
does not depend on this parameter. However, calculating
the matrix element of this operator in the effective theory and
studying its ultraviolet behaviour allows us to access the mass dependence
of the matrix element $\langle A | \bar{q} \Gamma Q | M (v) \rangle $.

The ultraviolet behaviour of the effective theory is investigated by
the renormalization group equations. Differentiating (\ref{expr})
with respect to the factorization scale $\mu$ yields the renormalization
group equation
\begin{equation}
\frac{d}{d \ln \mu} \left\{ Z \left( \frac{m_Q}{\mu} \right)
\langle A | \bar{q} \Gamma h_v | H (v) \rangle |_\mu \right\} = 0 
\end{equation}
from which we may obtain an equation which determines the
change of the coefficient $Z$ when the scale is changed
\begin{eqnarray} \label{rge}
&& \left( \frac{d}{d \ln \mu} + \gamma_J (\mu) \right) 
Z \left( \frac{m_Q}{\mu} \right)= 0
\\
&& \gamma_J (\mu) = \frac{d}{d \ln \mu} 
   \ln ( \langle A | \bar{q} \Gamma h_v | H (v) \rangle |_\mu )
\nonumber  .
\end{eqnarray}
The quantity $\gamma_J$ is called the anomalous dimension of the
operator $J = \bar{q} \Gamma h_v $ which is universal for all matrix elements of
this operator, since it is connected with the short distance behaviour of the
insertion of $J$. 

Eq.(\ref{rge}) describes the renormalization group scaling
in the effective theory. It allows to shift logarithms of the large
mass scale from the matrix element of $J$ into the 
coefficient $Z$: If the matrix element is renormalized at the large
scale $m_Q$ the logarithms of the type $\ln m_Q$ will apear in the matrix element
of $J$ while the coefficient $Z$ at this scale will simply 
be 
\begin{equation}
Z(1) = a_{00} + a_{10} \alpha_s (m_Q) + a_{20} \alpha_s^2 (m_Q) 
              + a_{30} \alpha_s^3 (m_Q) + \cdots
\end{equation}
The renormalization
group equation (\ref{rge}) allows to lower the renormalization point
from $m_Q$ to $\mu$; the matrix element renormalized at $\mu$ will
not contain any logarithms of $m_Q$ any more, they will appear in
the coefficient $Z$ in the way shown in (\ref{zexp}). 

In all cases relevant in the present context the matrix elements will 
be matrix elements involving
hadronic states, which are in most cases impossible to calculate from
first principles. However, eq.(\ref{rge}) allows to extract the
short distance piece, i.e.\ the logarithms of the large mass $m_Q$
and to separate it into the Wilson coefficients.

The anomalous dimension may be calculated in perturbation theory
in powers of the coupling constant $g$ of the theory.
In general, in a renormalizable theory the coupling constant depends
on the scale $\mu$ at which the theory is renormalized. The scale
dependence of the coupling constant is determined by the 
$\beta$ function
\begin{equation} \label{gml}
\frac{d}{d \ln \mu} g (\mu) = \beta (\mu) .
\end{equation}
In a mass independent scheme the renormalization
group functions $\gamma_{\cal O}$ and $\beta$ will depend on the
scale $\mu$ only through their dependence on the  coupling constant
\begin{equation}
\beta = \beta(g(\mu)) \qquad
\gamma_J = \gamma_J( g(\mu)) . 
\end{equation}
Hence we may rewrite the renormalization group equation (\ref{rge}) 
as 
\begin{equation} \label{rg1}
\left( \mu \frac{\partial}{\partial \mu}
+ \beta(g) \frac{\partial}{\partial g} +
\gamma_J (g) \right)  Z \left( \frac{m_Q}{\mu}, g \right) = 0 .
\end{equation}
The renormalization group functions $\beta$ and $\gamma_{\cal O}$
are calculated in perturbation theory; the first term of the $\beta$
function on QCD is obtained from a one-loop calculation and is given by
\begin{equation} \label{beta0}
\beta(g) = - \frac{1}{(4 \pi)^2} \left( 11 - \frac{2}{3} n_f \right)
           g^3 + \cdots ,
\end{equation}
where $n_f$ is the number of active flavors, i.e.\ the number
of flavors with a mass less than $m_Q$.

With this input the renormalization group equation may be solved to
yield
\begin{equation} \label{LLA}
Z \left( \frac{m_Q}{\mu} \right) = a_{00} \left( 
\frac{\alpha_s (\mu)}{\alpha_s (m_Q)} \right)^
          {\displaystyle -\frac{48 \pi^2}{33 - 2 n_f} \gamma_1}
\end{equation}
where $\gamma_1$ is the first coefficient in the perturbative expansion of
the anomalous dimension $\gamma_J = \gamma_1 g^2 + \cdots $ and 
$\alpha_s (\mu)$ is the one loop expression for the running coupling
constant of QCD
\begin{equation} \label{alpha}
\alpha_s (\mu) = \frac{12 \pi}{(33-2n_f) \ln(\mu^2/\Lambda_{QCD}^2)}
\end{equation}
which is obtained from solving (\ref{gml}) using (\ref{beta0}). 
This expression corresponds to a summation of the leading logarithms
$(\alpha_s \ln m_Q)^n$ which is  achieved by a one-loop calculation 
of the renormalization group functions $\beta$ and $\gamma_{\cal Q}$;  
in other words, in this way a resummation of the
first column of the expansion (\ref{zexp}) is obtained.

In a similar way one may also resum the second column of (\ref{zexp}), 
if the renormalization group functions $\beta$ and $\gamma$ are calculated 
to two loops and the finite terms of the one loop expression are included.

Finally, the case we have considered as an example is indeed very simple; 
in general all operators of a given dimension may mix under renormalization, 
i.e. instead of a simple anomalous dimension a matrix of anomalous dimensions
may occur. For more details on this I refer the reader to a textbook discussion 
of these issues as given e.g. in \cite{collins}. 

\subsection{Heavy Quark Symmetries}
The main impact of the heavy quark limit is due to two additional 
symmetries which are not present in full QCD; the first is a heavy flavour 
symmetry and the second one is the so-called spin symmetry. The 
presence of these symmetries implies Wigner-Eckart theorems for 
transition matrix elements which have far-reaching phenomenological consequences.

We shall first study the heavy flavour symmetry. The interaction of 
the quarks with the gluons is flavour independent; all flavour dependence 
in QCD is only due to the different quark masses. In the $1/m_Q$ expansion 
the leading order Lagrangian is mass independent and hence a flavour 
symmetry appears relating heavy quarks moving with the same velocity. 

For the case of two heavy flavours $b$ and $c$ one has to leading order the 
Lagrangian \cite{HQET}  
\begin{equation}
{\cal L}_{heavy} = \bar{b}_v (v \cdot D) b_v + \bar{c}_v (v \cdot D) c_v ,
\end{equation}
where $b_v$ ($c_v$) is the field operator $h_v$ for the $b$ ($c$) quark 
moving with velocity $v$ and $D = \partial + i g A$ is the QCD covariant 
derivative.   
This Lagrangian is obviously invariant under the $SU(2)_{HF}$ rotations  
\begin{equation}
\left( \begin{array}{c} b_v \\ c_v \end{array} \right) \to 
U_v \left( \begin{array}{c} b_v \\ c_v \end{array} \right) \quad 
U_v \in SU(2)_{HF} .
\end{equation}
We have put a subscript $v$ for the transformation matrix $U$, since 
this symmetry only relates heavy quarks moving with the same velocity.

The second symmetry  is the heavy-quark spin symmetry. As is clear form 
the Lagrangian in the heavy-mass limit, both spin degrees of freedom 
of the heavy quark couple in the same way to the gluons; we may 
rewrite the leading-order Lagrangian as 
\begin{equation}
{\cal L} = \bar{h}_v^{+s} (iv D) h_v^{+s} + \bar{h}_v^{-s} (iv D) h_v^{-s},
\end{equation}
where $h_v^{\pm s}$ are the projections of the heavy quark field on a 
definite spin direction $s$
\begin{equation}
h_v^{\pm s} = \frac{1}{2} (1 \pm \gamma_5 \fmslash{s}) h_v, 
\quad s\cdot v = 0 .
\end{equation}
This Lagrangian has a symmetry under the rotations of the heavy quark 
spin and hence all the heavy hadron states moving with the velocity $v$
fall into spin-symmetry doublets as $m_Q \to \infty$. In Hilbert space 
this symmetry is generated by operators $S_v (\epsilon)$ as
\begin{equation}
[ h_v , S_v (\epsilon) ] = i \fmslash{\epsilon} \fmslash{v} \gamma_5 h_v
\end{equation}
where $\epsilon$ with $\epsilon^2 = -1$ is the rotation axis.  
The simplest spin-symmetry doublet in the mesonic case consists of the 
pseudoscalar meson $H(v)$ and the corresponding vector meson 
$H^* (v,\epsilon)$, since a spin rotation yields
\begin{equation}
\exp\left(iS_v(\epsilon) \frac{\pi}{2} \right) | H (v) \rangle = 
(-i) | H^* (v,\epsilon) \rangle ,
\end{equation}
where we have chosen an arbitrary phase to be $(-i)$. 

In the heavy-mass limit the spin symmetry partners have to be 
degenerate and their splitting has to scale as $1/m_Q$. In other 
words, the quantity 
\begin{equation}
\lambda_2 = \frac{1}{4} (M_{H^*}^2 - M_H^2) 
\end{equation}
has to be the same for all spin symmetry doublets of heavy ground state 
mesons. This is well supported by data: For both the $(B,B^*)$ and the 
$(D,D^*)$ doublets one finds a value of $\lambda_2 \sim 0.12$ GeV${}^2$.
This shows that 
the spin-symmetry partners become degenerate in the infinite mass limit and 
the splitting between them scales as $1/m_Q$. 

In the infinite mass limit the symmetries imply relations between matrix 
elements involving heavy quarks.  For a transition between
heavy ground-state mesons $H$  (either pseudoscalar or vector)
with heavy flavour $f$ ($f'$) moving with velocities $v$ ($v'$), one
obtains in the heavy-quark limit
\begin{equation} \label{WET}
\langle H^{(f')} (v') | \bar{h}^{(f')}_{v'} \Gamma h^{(f)}_v
|  H^{(f)} (v) \rangle 
 = \xi (vv') \mbox{ Tr } 
\left\{ \overline{{\cal H} (v)} \Gamma {\cal H}(v) \right\} ,
\end{equation}
where $\Gamma$ is some arbitrary Dirac matrix and $H(v)$ are the 
representation matrices for the two possibilities of coupling 
the heavy quark spin to the spin of the light degrees of freedom, 
which are in a spin-1/2 state for ground state mesons   
\begin{equation} \label{mesonrep}
{\cal H}(v) = \frac{\sqrt{M_H}}{2} \left\{ \begin{array}{l l}
        (1+\fmslash{v}) \gamma_5 & 0^-, \, (\bar{q} Q)  \mbox{ meson} \\
        (1+\fmslash{v}) \fmslash{\epsilon}  & 1^- , \, (\bar{q} Q) \mbox{ meson} \\
                            & \mbox{with polarization } \epsilon .
       \end{array} \right.  
\end{equation}  
Due to the spin and flavour independence of the heavy mass limit 
the Isgur--Wise function $\xi$ is the only non-perturbative information 
needed to describe all heavy to heavy transitions within a spin-flavour 
symmetry multiplet.  

Excited mesons have been studied in \cite{Falk}. They may be classified 
by the angular momentum of the light degrees of freedom $j_l$, which 
is coupled with the heavy quark spin $S$ to the total angular momentum 
$J$ of the meson. Furthermore, the orbital angular momentum $\ell$ determines 
the parity $P = (-1)^{\ell+1}$ of the meson. For a given $\ell> 0$ we can have 
$j_l = \ell \pm 1/2$ and the coupling of the heavy quark spin yields 
two spin symmetry doublets $(J = \ell-1 , J = \ell)$ and 
$(J = \ell, J = \ell +1)$. For example, the lowest positive parity 
$\ell = 1$ mesons are two spin symmetry doublets $(0^+, 1^+)$ and 
$(1^+, 2^+)$. In the $D$ meson system these states have been observed
\cite{CLEOD} and behave as predicted by heavy quark symmetry \cite{EHQexcited}

Similarly as for the mesons heavy-quark symmetries imply that
only one form factor is needed to describe heavy to heavy transitions within 
a spin flavour symmetry multiplet; in other words, there is an Isgur Wise 
function for each multiplet.   

The ground state baryons have been studied in \cite{IWbary,Geobary,MRRbary}. 
According to the particle data group they are classified as follows
\begin{eqnarray}
\label{Lam} 
\Lambda_h &=& [(qq^\prime)_0 h]_{1/2}  \qquad \Xi_h^\prime = [(qs)_0 h]_{1/2} \\  
\label{Sig}
\Sigma_h &=& [(qq^\prime)_1 h]_{1/2}  \qquad \Xi_h = [(qs)_1 h]_{1/2} 
\qquad \Omega_h = [(ss)_1 h]_{1/2} \\ 
\label{SigStar}
\Sigma_h^* &=& [(qq^\prime)_1 h]_{3/2} \qquad 
\Xi_h^* = [(qs)_1 h]_{3/2} \qquad  
\Omega_h^* = [(ss)_1 h]_{3/2}.  
\end{eqnarray}
Here, $q, q^\prime$ refer to $u$ and $d$ quarks, $q \neq q^\prime$ for the
$\Lambda_h$, but $q$ may be the same as $q^\prime$ for the $\Sigma_h$ and
$\Sigma_h^*$. The first subscript (0, 1) is the total spin of the light
degrees of freedom, while the second subscript (1/2, 3/2) is the total spin 
of the baryon. 

Spin symmetry forces these baryons into spin symmetry doublets. For the 
$\Lambda$-type baryons (\ref{Lam}) the spin rotations are simply a subset of the Lorentz
transformations, since the light degrees of freedom are in a spin-0 state. 
The corresponding spin symmetry doublet is in this case given by the two 
polarization directions of the heavy baryon.    
From the point of view of heavy quark symmetries the $\Lambda$-type baryons
are the simplest hadrons, although from the quark model 
point of view they are composed of three quarks. 

The baryons with the light degrees of freedom in a spin one state may be 
represented by a pseudovector-spinor object $R^\mu$ with $v_\mu R^\mu=0$
\footnote{One could as well represent the light degrees of freedom by an 
          antisymmetric tensor instead of a pseudovector; this is a completely
          equivalent formulation \cite{MRRbary}.}.  
In general $\gamma_\mu R^\mu \neq 0$ because $R^\mu$ contains 
spin 1/2 contributions as well as spin 3/2 parts. 
In other words, $R^\mu$ contains a Rarita-Schwinger field as well as
a Dirac field. Under Lorentz transformations $R^\mu$ behaves as 
\begin{equation}
R^\mu(v) \rightarrow {\Lambda^\mu}_\nu  D(\Lambda) R^\nu(\Lambda v),
\end{equation}
where $\Lambda_{\mu \nu}$ and $D(\Lambda)$ are the Lorentz transformations 
in the vector and spinor representation respectively, while 
under spin rotations we have
\begin{equation} \label{sigspin}
R^\mu(v) \rightarrow - \gamma_5 \fmslash{v} \fmslash{\epsilon} R^\mu(v) .
\end{equation}

The spin-3/2 component of the the pseudovector-spinor object 
corresponding to the $\Sigma^*_h$ is projected 
out by contracting with $\gamma_\mu$
\begin{equation}
\gamma_\mu R^\mu _{\Sigma_h^*} = 0 .
\end{equation}
The rest of the independent components of $R$ correspond to $\Sigma_h$ baryon:
\begin{equation}
R^\mu_{\Sigma_h} = {1 \over \sqrt{3}} (\gamma^\mu + v^\mu) \gamma_5
u_{\Sigma_h},
\end{equation}
where $u_{\Sigma_h}$ is the Dirac spinor of the $\Sigma_h$ state. Similar 
expressions hold for the non-strange baryons $\Xi^{(*)}_h$ and $\Omega_h^{(*)}$. 

The spin rotation (\ref{sigspin})  transform the $\Sigma$-like baryons into 
the $\Sigma^*$ states 
and vice versa. Thus the spin symmetry doublets for the ground state baryons 
are given by the two polarization directions of the baryons in (\ref{Lam}), and
by the two states with corresponding light quark flavour numbers in (\ref{Sig}) 
and (\ref{SigStar}). 

Similar to the case of mesons one may derive a Wigner-Eckart theorem for the 
spin symmetry doublets of the baryons 
\begin{equation}\label{lambda} 
\langle \Lambda_{h}(v) | \bar{h}\Gamma h' | \Lambda_{h'}(v') \rangle = A(v\cdot v')
\bar{u}_{\xi_h}(v) \Gamma u_{\xi_{h'}}(v'),
\end{equation}
where we have allowed for the possibility of two heavy quark flavours $h$
and $h'$. In the same way, one obtains two form factors for the 
$\Sigma^{(*)}_h \to \Sigma^{(*)}_{h'}$.   
\begin{eqnarray} \label{ss1}
&& \langle \Sigma^{(*)}_h (v) | \bar{h}_v \Gamma h_{v'} |\Sigma^{(*)}_h \rangle \\
&& = \bar{R}^{\mu}_{\Sigma_h^{(*)}} (v) \Gamma {R}^{\nu}_{\Sigma_h^{(*)}}(v')
\, \left[ B(v\cdot v')g_{\mu \nu} + C(v\cdot v')v'_\mu v_\nu \right] .
\nonumber
\end{eqnarray}
Finally, parity does not allow for transitions between $\Lambda$ and 
$\Sigma^{(*)}$ type baryons  
\begin{equation}\label{sl1}
\langle \Sigma^{(*)}_h(v) | \bar{h}_v \Gamma h_{v'} |\Lambda_h (v) \rangle = 0 , 
\end{equation}
and hence these transitions are not only suppressed by the flavour symmetry 
of the light degrees of freedom, but additionally by heavy quark symmetry.

Excited baryons my be studied along the same lines as for the mesons. 
The spin symmetry doublets as well as the restrictions on transition matrix 
elements have been studied in \cite{Falk}.  

Heavy quark symmetries thus lead to a strong reduction of the number of 
independent from factors that describe current induced transitions among heavy 
hadrons. In addition to that the symmetries even allow us to obtain the 
normalization of some of these form factors. Since the currents 
\begin{equation}
J^{hh'} = \bar{h}_v \gamma_\mu h^\prime_v = v_\mu \bar{h}_v h^\prime_v     
\end{equation}
are the generators of heavy flavour symmetry in the velocity sector $v$, 
the normalization of the Wigner-Eckard theorems 
(\ref{WET},\ref{lambda},\ref{ss1}) is known at the non-recoil point $v=v'$.  
By standard arguments one obtains for the mesons 
\begin{equation} \label{normxi}
\xi (vv' = 1) = 1 ,
\end{equation}
while the corresponding relation for the baryons is 
\begin{eqnarray}
A(vv' = 1) &=& \sqrt{m_{\Lambda_h} m_{\Lambda_{h'}}} \label{normA}\\
B(vv' = 1) &=& \sqrt{m_{\Sigma^{(*)}_h} m_{\Sigma^{(*)}_{h'}}} , \label{NormB} 
\end{eqnarray}
where the factor involving the square root of 
the masses  means that the hadron states
in (\ref{lambda}) are normalized relativistically. 

Up to now we have considered only the consequences of heavy quark symmetries 
for the leading terms of the $1/m_Q$ expansion. However, the additional 
symmetries also restrict the subleading terms and one of these restrictions 
is called Lukes theorem \cite{Lu91}. It is a 
generalization of 
the Ademollo Gatto theorem \cite{AdemolloGatto}, which states that in 
the presence of explicit 
symmetry breaking the matrix elements of the currents that generate the 
symmetry are still normalized up to terms which are second order in the 
symmetry breaking interaction. 

For the case at hand the relevant symmetry is the heavy flavor symmetry. 
This symmetry is an $SU(2)$ symmetry and is generated by three operators
$Q_\pm$ and $Q_3$ with 
\begin{eqnarray}
&& Q_+ = \int d^3 x \, \bar{b}_v (x) c_v (x) \quad 
Q_- = \int d^3 x \, \bar{c}_v (x) b_v (x)   \nonumber \\ 
&& Q_3 = \int d^3 x \, (\bar{b}_v (x) b_v (x) - \bar{c}_v (x) c_v (x)) 
\\ \nonumber  
&& [Q_+ , Q_- ] = Q_3 \qquad [Q_+ , Q_3 ] = -2 Q_+ \qquad (Q_+)^\dagger = Q_-
\end{eqnarray}

Let us denote the ground state flavour symmetry 
multiplet as $| B \rangle$ and $| D \rangle$. Then the operators act in 
the following way
\begin{eqnarray}
&& Q_3 | B \rangle = | B \rangle \qquad Q_3 | D \rangle = - | D \rangle \nonumber \\      
&& Q_+ | D \rangle = | B \rangle \qquad Q_- | B \rangle =   | D \rangle \\
&& Q_+ | B \rangle = Q_- | D \rangle = 0 \nonumber . 
\end{eqnarray}
The Hamiltonian of this system has a $1/m_Q$ expansion of the form 
\begin{eqnarray}
H &=& H_0^{(b)} + H_0^{(c)} + 
    \frac{1}{2m_b} H_1^{(b)} + \frac{1}{2m_c} H_1^{(c)} + \cdots \\
\nonumber
  &=& H_0^{(b)} + H_0^{(c)} + 
    \frac{1}{2} \left(\frac{1}{2m_b}+\frac{1}{2m_c}\right) 
                (H_1^{(b)}+H_1^{(c)})  \\  
\nonumber
 && \hphantom{H_0^{(b)} + H_0^{(c)}}
+ \frac{1}{2} \left(\frac{1}{2m_b}-\frac{1}{2m_c}\right) 
                (H_1^{(b)}-H_1^{(c)}) + \cdots \\
&=& H_{symm} + H_{break} \nonumber .
\end{eqnarray}  
In the second equation, the first line is still symmetric under heavy flavour 
$SU(2)$ while the term in the second line does not commute any more with $Q_\pm$, 
but it still commutes with $Q_3$. In other words, to order $1/m_Q$ we still 
have common eigenstates of $H$ and $Q_3$, which we shall denote as 
$\tilde{| B \rangle}$ and $\tilde{| D \rangle}$. Sandwiching the 
commutation relation we get 
\begin{eqnarray}
1&=& \tilde{\langle B |} Q_3 \tilde{| B \rangle} = 
   \tilde{\langle B |} [Q_+ , Q_- ] \tilde{| B \rangle}  
\\ \nonumber 
 &=& \sum_n \left[ \tilde{\langle B |} Q_+ \tilde{| n \rangle}
     \tilde{\langle n |}  Q_-  \tilde{| B \rangle} - 
     \tilde{\langle B |} Q_- \tilde{| n \rangle}
     \tilde{\langle n |}  Q_+  \tilde{| B \rangle}\right] \\
\nonumber
&=& \sum_n \left[ |\tilde{\langle B |} Q_+ \tilde{| n \rangle}|^2 -
                  |\tilde{\langle B |} Q_- \tilde{| n \rangle}|^2 \right]
\end{eqnarray}
where $\tilde{| n \rangle}$ form a complete set of states of the Hamiltonian 
$H_{symm} + H_{break}$. The matrix elements may be written as 
\begin{equation}
\tilde{\langle B |} Q_\pm \tilde{| n \rangle} = 
\frac{1}{E_B - E_n} \tilde{\langle B |} [H_{break}, Q_\pm ] \tilde{| n \rangle}
\end{equation}
where $E_B$ and $E_n$ are the energies of the states $\tilde{| B \rangle}$
and $\tilde{| n \rangle}$ respectively. In the case 
$\tilde{| n \rangle} = \tilde{| D \rangle}$ the matrix element will 
be of order unity, since both the numerator as well as the energy difference in the 
denominator are of the order of the symmetry breaking. For all other states the 
energy difference in the denominator is non-vanishing in the symmetry limit, 
and hence this difference is of order unity; thus the matrix element for these 
states will be of the order of the symmetry breaking. From this we conclude
\begin{equation}
\tilde{\langle B |} Q_+ \tilde{| D \rangle} = 
1 + {\cal O}\left[\left(\frac{1}{2m_b}-\frac{1}{2m_c}\right)^2\right] .
\end{equation}
In particular, the weak transition currents at the non-recoil point $v=v'$
are proportional to these symmetry generators and hence we may conclude that
for some of these matrix elements we only have corrections of the order
$1/m_Q^2$. 

Another restriction on the $1/m_Q$ expansion is imposed by the so-called 
reparametrization invariance \cite{repara} which is basically the remnant 
of the original Lorentz covariance of full QCD.  
The full theory depends only on the 
momentum of 
the heavy quark $P$, and the splitting of this momentum in an on-shell 
part $m_Q v$ and a residual momentum $k$ corresponding to the covariant 
derivative acting on the heavy static field $h_v$ is arbitrary. Formally 
this means that the Lagrangian does not depend on the velocity $v$, if 
all orders of the $1/m_Q$ expansion are included; the $v$ dependence only 
enters once the expansion is truncated. 

Using the representation (\ref{HarvLag}) and (\ref{HarvField}) the 
Lagrangian is invariant under the transformation 
\begin{eqnarray} \label{repara}
&& v \to v + \delta v \qquad v \cdot \delta v = 0 \\
&& h_v \to h_v + \frac{\delta v \hskip -0.75em / \hskip 0.25em }{2} 
\left(1+ P_- \frac{1}{2m + ivD} i \fmslash{D} \right) h_v \nonumber \\
&& iD \to - m \, \delta v \nonumber .
\end{eqnarray}
This invariance is the so-called reparametrization invariance, 
which has non-trivial consequences, since it relates terms of different 
orders of the $1/m$ expansion.

\section{Application to Exclusive Decays}
The heavy mass limit and the resulting additional symmetries allow 
us to restrict the matrix elements which occur in weak transitions 
of heavy hadrons. We shall consider in the following in some detail 
the semileptonic $b \to c$ transition, which we shall treat as a heavy 
to heavy decay. In section 3.2 we investigate the consequences of 
heavy quark symmetries for transitions of the heavy to light type.

\subsection{Transitions of the type Heavy $\to$ Heavy}
For the case of a heavy to heavy transition the 
Wigner Eckart theorem (\ref{WET}) implies that there is only a single 
form factor which describe the weak decays of heavy 
hadrons; furthermore, the heavy mass limit yields the normalization 
of this form factor at the kinematic point $v = v'$. 

Treating both the $b$ and the $c$ quark as heavy, the 
semileptonic decays $B \to D^{(*)} \ell \nu$ are the phenomenologically 
relevant examples. The matrix elements for these transitions are in general 
parametrized in terms of six form factors
\begin{eqnarray}
 \langle D (v') | \bar{c} \gamma_\mu b | B(v) \rangle  &=&  \sqrt{m_B m_D}
 \left[ \xi_+ (y) (v_\mu + v'_\mu)
     + \xi_- (y) (v_\mu - v'_\mu) \right] \\ 
 \langle D^* (v',\epsilon) | \bar{c} \gamma_\mu b | B(v) \rangle  &=& 
       i \sqrt{m_B m_{D^*}} 
\xi_V (y) \varepsilon_{\mu \alpha \beta \rho} \epsilon^{*\alpha}
                       v^{\prime \beta} v^\rho  \\ 
\langle D^* (v',\epsilon) | \bar{c} \gamma_\mu \gamma_5 b | B(v) \rangle 
   &=& \sqrt{m_B m_{D^*}}
       \left[ \xi_{A1} (y) (vv'+1) \epsilon^*_\mu
      -  \xi_{A2} (y) (\epsilon^* v)  v_\mu \right. \nonumber \\
   && \qquad \qquad \left.   -  \xi_{A2} (y) (\epsilon^* v)  v'_\mu \right] ,
\end{eqnarray} 
where we have defined $y = vv'$. Due to the Wigner Eckart theorem 
(\ref{WET}) these six from factors are related to the Isgur Wise 
function by  
\begin{equation}
\xi_i (y) = \xi (y) \mbox{ for } i = +,V,A1,A3, \qquad
\xi_i (y) = 0       \mbox{ for } i = -,A2 .
\end{equation} 
Since heavy quark symmetries also yield the normalization of the 
Isgur Wise function, we know the absolute value of the differential 
rate at the point $v = v'$ in terms of the meson masses and $V_{cb}$. 
Hence we may use this to extract $V_{cb}$ from these decays in a model 
independent way 
 by extrapolating 
the lepton spectrum to the kinematic endpoint $v=v'$. 
Using the mode $B \to D^{(*)} \ell \nu$ one obtains the relation
\begin{equation} \label{extra}
\lim_{v \to v'} \frac{1}{\sqrt{(vv')^2-1}} \frac{d \Gamma}{d(vv')} = 
\frac{G_F^2}{4 \pi^3} |V_{cb}|^2 (m_B - m_{D^*})^2 m_{D^*}^3 
|\xi_{A1} (1)|^2 , 
\end{equation}
where $\xi_{A1}$ is equal to the Isgur Wise function in the 
heavy mass limit, and hence $\xi_{A1} (1) = 1$.  

Corrections to this relation have been calculated along the lines 
outlined above in leading and subleading order. A complete discussion 
may be found in more extensive review articles (see e.g. Neubert's review 
\cite{reviews}), including  reference to the original papers. Here we only 
state the final result
\begin{eqnarray}
&& \xi_{A1} (1)  =  x^{6/25} \left[ 1 + 1.561 \frac{\alpha_s (m_c) - \alpha_s (m_b)}{\pi}
              - \frac{8 \alpha_s (m_c)}{3 \pi} \right.  
\\ \nonumber
&& \quad + z \, \left\{\frac{25}{54} - \frac{14}{27} x^{-9/25} 
              + \frac{1}{18} x^{-12/25} + \frac{8}{25} \ln x 
       \right\} 
\left. - \frac{\alpha_s (\bar{m})}{\pi} \frac{z^2}{1-z} \ln z \right] 
 + \, \delta_{1/m^2}  ,
\label{msquared}
\end{eqnarray}
where we use the abbreviations
$$ 
x = \frac{\alpha_s (m_c)}{ \alpha_s (m_b)}, \quad z = \frac{m_c}{m_b}
$$ 
and $\bar{m}$ is a scale somewhere between $m_b$ and $m_c$.

Up to the term $\delta_{1/m^2}$ all these contributions may be calculated 
perturbatively, including the dependence on $z$. The quantity 
$\delta_{1/m^2}$ parametrizes the non-perturbative contributions, which enter 
here at order $1/m^2$. These corrections may be expressed in terms of the 
kinetic energy $\lambda_1$, the chromomagnetic moment $\lambda_2$, which are
given in (\ref{lam1}) and (\ref{lam2}) respectively, and matrix elements 
involving time-ordered products between the current and the corrections of the 
Lagrangian  
\begin{eqnarray} 
\delta_{1/m^2}  &=&  - \left(\frac{1}{2m_c} \right)^2
   \frac{1}{2} \left( -\lambda_1 + \lambda_2  \vphantom{\int} \right.
\\ \nonumber  &+&   \left. 
(-i)^2 \frac{1}{2\sqrt{M_B M_D}} \int d^4 x \,d^4y \,
\langle B^*(v,\epsilon) | T \left[ {\cal L}_b^{(1)} (x) \bar{b}_v c_v 
                                      {\cal L}_c^{(1)} (y) \right]
                   | D^*(v,\epsilon) \rangle \right) \\
&+& \vphantom{\int}{\cal O} (1/m_c^3,1/m_b^2, 1/(m_c m_b) ) ,  \nonumber
\end{eqnarray}
where ${\cal L}_Q^{(1)}$ is the first order Lagangian for the quark $Q$ as given 
in (\ref{HarvLag}) or (\ref{FWlag}) and $M_B$ ($M_D$) are the masses of the $B$ 
($D$) meson in the heavy quark limit.
Here we display only the largest contribution of order $1/m_c^2$; the complete 
expression, including the $1/m_b^2$ and $1/(m_c m_b)$ terms,  may be found 
in \cite{FN92,zerec}. 

Thus the correction $\delta_{1/m^2}$ is given in terms of $\lambda_1$ defined in 
(\ref{lam1}), $\lambda_2$ given in (\ref{lam2}) and a non-local matrix element
involving a time-ordered product. The problem concerning the determination of 
$\lambda_1$ has been considered already above; 
similarly it is not easy to obtain information on the matrix element involving 
the time-ordered product, and thus the corrections of order $1/m^2$ will finally 
limit our ability to determine the CKM matrix element $V_{cb}$ in a model independent 
way, at least using the approach described here. 

Various estimates for $\delta_{1/m^2}$ have been given in the literature. The 
first estimate of this correction has been given in \cite{FN92} using the GISW
model \cite{GISW}, which is based on a wave function for the light quark. In this 
work $ \delta_{m^2} = -2\% \ldots -3\% $ has been obtained. Another estimate with 
weaker assumptions yields $\delta_{m^2} =  0   \ldots -5\%$ \cite{zerec}, but both 
estimates have been criticized recently as being too small. Based on heavy flavour 
sum rules it has been argued in \cite{SU94} that the $1/m^2$ corrections can be 
quite large $\delta_{m^2} =  0\% \ldots -8\% $ \cite{SU94}. These various estimates 
indicate the size of the theoretical error involved in the determination of $V_{cb}$
from the exclusive channel $B \to D^* \ell \bar{\nu}_\ell$; a generally accepted 
value for these corrections has been given recently \cite{Neubvcb}
\begin{equation}
\delta_{m^2} = -0.055 \pm 0.025
\end{equation}
from which one obtains 
\begin{equation} \label{xia1}
\xi_{A1} (1)  = 0.91 \pm 0.03
\end{equation}

This result has been used to extract $V_{cb}$ from CLEO \cite{CLEOvcb} 
as well as from LEP data \cite{ALEPHvcb}. 
The values obtained are 
\begin{eqnarray}
|V_{cb}| &=& 0.0386 \pm 0.0019 \pm 0.0020 \pm 0.0014 \qquad \mbox{CLEO} \\
|V_{cb}| &=& 0.0392 \pm 0.0025 \pm 0.0027 \pm 0.0015 \qquad \mbox{ALEPH} 
\end{eqnarray} 
where (\ref{xia1}) has been used.
Note that the third error in $|V_{cb}|$ 
is due to the theoretical uncertainties, which by now almost match the 
experimental ones.

\subsection{Transitions of the type Heavy $\to$ Light}
Heavy quark symmetries may also be used to restrict the independent
form factors appearing in heavy to light decays. For the decays of heavy 
mesons into light $0^-$ and $1^-$ particles heavy quark symmetries restrict 
the number of independent form factors to six, which is just the 
number needed to parametrize the semileptonic decays of this type. 
Furthermore, no absolute normalization of form factors may be obtained from 
heavy quark symmetries in the heavy to light case; only the relative 
normalization of $B$ meson decays heavy to light transitions may be obtained 
from the corresponding $D$ decays.

In general we shall discuss matrix elements of a heavy to light
current which have the following structure
\begin{equation} \label{htlg}
J = \langle  A | \bar{\ell} \Gamma h_v | H (v) \rangle ,
\end{equation}
where $\Gamma$ is an arbitrary Dirac matrix,
$\ell$ is a light quark
($u$, $d$ or $s$) and
$A$ is a state involving only light degrees of freedom.

Spin symmetry implies that the heavy quark index hooks directly the
to the heavy quark index of the Dirac matrix of the current. Thus one
may write for the transition matrix element~(\ref{htlps})
\begin{equation} \label{htltrg}
\langle  A | \bar{\ell} \Gamma h_v | H (v) \rangle =
\mbox{ Tr } \left( {\cal M}_A \Gamma H(v) \right)
\end{equation}
where the matrix $H(v)$ representing the heavy meson has been
given in (\ref{mesonrep}).
The matrix ${\cal M}_A$ describes the light degrees of freedom
and is the most general matrix which may
be formed from the kinematical variables involved. Furthermore, if
the energies of the particles in the state $A$ are small, i.e.\ of
the order of $\Lambda_{QCD}$, the matrix ${\cal M}_A$ does not depend
on the heavy quark; in particular it does not depend on the heavy mass
$m_H$.
In the following we
shall discuss some examples.

The first example is the heavy meson decay constant, where the
state $A$ is simply the vacuum state. The heavy meson decay
constant is defined by
\begin{equation}
\langle 0 | \bar{\ell} \gamma_\mu \gamma_5 h_v | H (v) \rangle =
f_H m_H v_\mu , 
\end{equation}
and since $| A \rangle = | 0 \rangle$ the matrix ${\cal M}_0$ is simply the 
unit matrix times a dimensionful constant\footnote{%
  Note that contributions proportional to
  $\fmslash{v}$ may be eliminated using
  $$
  H(v) \fmslash{v} = - H(v).
  $$ }
and one has, using
(\ref{htltrg})
\begin{equation}
\langle 0 | \bar{\ell} \gamma_\mu \gamma_5 h_v | H (v) \rangle = \kappa
\mbox{ Tr } \left( \gamma \gamma_5 H(v) \right) = 2 \kappa
\sqrt{m_H} v_\mu .
\end{equation}
As discussed above the constant $\kappa$ does not depend on the
heavy mass and thus one infers the well-known
scaling law for the heavy meson
decay constant from the last two equation
\begin{equation} \label{sca}
f_H  \propto \frac{1}{\sqrt{m_H}}
\end{equation}
Including the leading and subleading QCD radiative corrections
one obtains a relation between $f_B$ and $f_D$
\begin{equation}
f_B  = \sqrt{ \frac{m_c}{m_b}}
\left( \frac{\alpha_s (m_b)}{\alpha_s (m_c)} \right)^{-6/25} 
\left[ 1 + 0.894 \frac{\alpha_s (m_c)-\alpha_s (m_b)}{\pi} \right]
f_D \sim 0.69 f_D .
\end{equation}

The second example are transitions of a heavy meson
into a light pseudoscalar meson, which we shall denote as $\pi$.
The matrix element corresponding to (\ref{htlg}) is
\begin{equation} \label{htlps}
J_P = \langle \pi(p) | \bar{\ell} \Gamma h_v | H (v) \rangle , 
\end{equation}
where $p$ is the momentum of the light quark,

The Dirac marix ${\cal M}_P$ for the light degrees of freedom
appearing now in (\ref{htltrg})
depends on $p$ and $v$.
It may be expanded in terms of the sixteen independent
Dirac matrices $1$, $\gamma_5$,  $\gamma_\mu$, $\gamma_5 \gamma_\mu$,
and $\sigma_{\mu \nu}$ taking into account that it has to behave
like a pseudoscalar.
The form factors appearing in the decomposition of
${\cal M}_P$
depend on the variable $v \cdot p$, the energy of the light meson
in the rest frame of the heavy one. In order to compare different
heavy to light transition by employing heavy flavor symmetry this
energy must be sufficiently small, since the typical scale for the
light degrees of freedom has to be of the order of $\Lambda_{QCD}$
to apply heavy quark symmetry\footnote{%
         Note that in this case the variable $v \cdot p$ ranges between
         $0$ and $m_H / 2$ where we have neglected the pion mass. Thus 
         at the upper end of phase space the variable $v \cdot p$ scales
         with the heavy mass and heavy quark symmetries are not applicable
         any more.}.  
For the case of a light pseudoscalar
meson the most general decomposition of ${\cal M}_P$ is
\begin{equation}
{\cal M}_P = \sqrt{v \cdot p} A (\eta ) \gamma_5
         + \frac{1}{\sqrt{v \cdot p}} B (\eta ) \gamma_5 \fmslash{p} ,
\end{equation}
where we have defined the dimensionless variable
\begin{equation} \label{eta}
\eta = \frac{v \cdot p}{\Lambda_{QCD}} .
\end{equation}

The form factors $A$ and $B$ are universal in the kinematic range
of small energy of the light meson, i.e.\ where the momentum
transfer to the light degrees of freedom is of the order $\Lambda_{QCD}$;
in this region $\eta$ is of order unity. This universality of the
form factors may be used to relate various kinds of heavy to
light transitions, e.g.\ the semileptonic decays like
$D \to \pi e \nu$, $D \to K e \nu$ or $B \to \pi e \nu$
and also the rare decays like $B \to K \ell^+ \ell^-$
or $B \to \pi \ell^+ \ell^-$ where $\ell$ denotes an electron or a muon.

As an example we give the relations between exclusive semileptonic
heavy to light decays. The relvant hadronic
current for this case may be expressed in terms of two form
factors
\begin{eqnarray} \label{frmfps}
\langle \pi(p) | \bar{\ell} \gamma (1-\gamma_5) h_v | H (v) \rangle
&=& F_1 (v \cdot p ) m_H v_\mu +
    F_2 (v \cdot p ) p_\mu
\\
&=& F_+ (v \cdot p ) (m_H v_\mu + p_\mu ) +
    F_- (v \cdot p ) q_\mu
\nonumber
\end{eqnarray}
where
\begin{equation}
F_\pm (v \cdot p ) = \frac{1}{2} \left( F_1 (v \cdot p ) \pm
                     F_2 (v \cdot p ) \right)
\end{equation}
Inserting this into (\ref{htlps}) one may express $F_\pm$ in terms
of the universal form factors $A$ and $B$
\begin{eqnarray} \label{f1hql}
F_1 (v \cdot p ) &=& F_+ (v \cdot p ) + F_- (v \cdot p )
=   -2 \sqrt{\frac{v \cdot p}{m_H}} A (\eta)
\\  \label{f2hql}
F_2 (v \cdot p ) &=& F_+ (v \cdot p ) - F_- (v \cdot p )
=   -2 \sqrt{\frac{m_H}{v \cdot p}} B (\eta)
\end{eqnarray}
From these relations one may read off the scaling of the form
factors with the heavy mass which was already
derived in \cite{iwlight}.

This may be used to normalize the semileptonic $B$ decays into light mesons
relative to the semileptonic $D$ decays. One obtains
\begin{equation} \label{bdphql}
F_\pm^B (v \cdot p ) =
\frac{1}{2} \left(\sqrt{\frac{m_D}{m_B}}
                  \pm \sqrt{\frac{m_B}{m_D}} \right)
                       F_+^D (v \cdot p )
+ \frac{1}{2} \left(\sqrt{\frac{m_D}{m_B}}
                  \mp \sqrt{\frac{m_B}{m_D}} \right)
                       F_-^D (v \cdot p )
\end{equation}
Note that $F_+$ for the $B$ decay is expressed in terms of
$F_+$ {\it and} $F_-$ for the $D$ decays. In the limit of vanishing
fermion masses only $F_+$ contributes, which means that the $F_-$ 
contribution to the rate is of the
order of $m_{lepton} / m_H$. Thus it will be extremely difficult to
determine experimentally.

The case of a heavy meson decaying into a light vector meson may be
treated similarly.
The matrix element for the transition of a heavy meson into a light
vector meson (denoted generically as $\rho$ in the following)
is given again by (\ref{htlg}) and is in this case
\begin{equation} \label{htlv}
J_V = \langle \rho(p,\epsilon) | \bar{\ell} \Gamma h_v | H (v) \rangle .
\end{equation}
Using (\ref{htltrg}) one has
\begin{equation} \label{htlvtr}
\langle \rho(p,\epsilon) | \bar{\ell} \Gamma h_v | H (v) \rangle =
\mbox{ Tr } \left( {\cal M}_V \Gamma H(v) \right) ,
\end{equation}
where now the Dirac matrix ${\cal M}_V$ has to be a linear
function of the polarization of the light vector meson.

The most general decomposition
is given in terms of four
dimensionless form factors
\begin{equation} \label{hqlv}
{\cal M}_V = \sqrt{v \cdot p} C(\eta) (v \cdot \epsilon)
    +  \frac{1}{\sqrt{v \cdot p}} D(\eta) (v \cdot \epsilon) \fmslash{p}
 + \sqrt{v \cdot p} E(\eta) \fmslash{\epsilon}
 +  \frac{1}{\sqrt{v \cdot p}} F(\eta) \fmslash{p} \fmslash{\epsilon}
\end{equation}
where the variable $\eta$ has been defined in (\ref{eta}).

Similar to the case of the decays into a light pseudoscalar meson
(\ref{htlvtr}) may be used to relate various exclusive heavy to
light processes in the kinematic range where the energy of the
outgoing vector meson is small. For example,
the semileptonic decays
$D \to \rho e \nu$, $D \to K^* e \nu$ and $B \to \rho e \nu$ are related
among themselves and all of them may be related to the rare heavy
to light decays $B \to K^* \ell^+ \ell^-$
and $B \to \rho \ell^+ \ell^-$
with $\ell = e, \mu$.

Data on these decays are still very sparse; there are first measurements 
of the decays $B \to \pi \ell \nu$ and  $B \to \rho \ell \nu$ from CLEO
\cite{Berkel}, from which total rates may be obtained. From this one 
may extract a value of $V_{ub}$ by employing form factor models, and the 
value given by CLEO is 
\begin{equation} 
|V_{ub}| = (3.3 \pm 0.2 ^{+0.3}_{-0.4} \pm 0.7) \times 10^{-3}
\end{equation}
where the last uncertainty represents the variation of the result 
between different models. In order to perform a model independent determination 
along the lines discussed above a good measurement of the lepton energy 
spectra in these decays is needed. 

Finally we comment on the heavy to light transitions of baryons. For the 
$\Lambda$-type heavy baryons (\ref{Lam}) spin symmetry relates different
polarizations of the same particle and thus imposes interesting
constraints. Consider for example the matrix element of an operator
$\bar{\ell} \Gamma h_v$ between a heavy $\Lambda_Q$ and a light
spin-1/2 baryon $B_\ell$. It is described by only two form
factors,
\begin{equation}
\langle B_\ell (p) | \bar{\ell} \Gamma h_v | \Lambda_Q  (v) \rangle =
\bar{u}_\ell (p) \{ F_1 (v \cdot p ) + \fmslash{v} F_2 (v \cdot p) \} \Gamma
u_{\Lambda_Q} (v) .
\end{equation}
Thus in this particular case spin symmetry greatly reduces the
number of independent Lorentz-invariant amplitudes which 
describe the heavy to light transitions.

This has some interesting implications for exclusive semileptonic 
$\Lambda_c$ decays. For the case of a left handed current 
$\Gamma = \gamma_\mu (1-\gamma_5)$, the semileptonic decay 
$\Lambda_c \to \Lambda \ell \bar{\nu}_\ell$ is in 
general parametrized in terms  of six form factors
\begin{eqnarray}
\langle \Lambda (p) | \bar{q} \gamma_\mu (1-\gamma_5) c | \Lambda_c (v) \rangle &=&
\bar{u} (p) \left[ f_1 \gamma_\mu
               + i f_2 \sigma_{\mu \nu} q^\nu
               +   f_3 q^\mu  \right] u(p')  \nonumber \\
&+& \bar{u} (p) \left[ g_1 \gamma_\mu
               + i g_2 \sigma_{\mu \nu} q^\nu
               +   g_3 q^\mu  \right] \gamma_5 u(p') ,
\end{eqnarray}
where $p' = m_{\Lambda_c} v$ is the momentum of the $\Lambda_c$ whereas
$q = m_{\Lambda_c} v - p$ is the momentum transfer. From this one
defines the ratio $G_A / G_V$ by
\begin{equation}
\frac{G_A}{G_V} = \frac{g_1 (q^2 = 0)}{f_1 (q^2 = 0)} .
\end{equation}

In the heavy $c$ quark limit one may relate the six form factors
$f_i$ and $g_i$ ($i=1,2,3$) to the two form factors $F_j$
($j=1,2$)
\begin{eqnarray}
f_1 &=& - g_1 = F_1 + \frac{m_\Lambda}{m_{\Lambda_c}} F_2 \\
f_2 &=& f_3 = -g_2 = -g_3 = \frac{1}{m_{\Lambda_c}} F_2
\end{eqnarray}
from which one reads off $G_A / G_V = -1$. This ratio is acessible
by measuring in semileptonic decays $\Lambda_c \to \lambda \ell \bar{\nu}_\ell$ 
the polarization variable $\alpha$
\begin{equation}
\alpha = \frac{2 G_A G_V}{G_A^2 + G_V^2} 
\end{equation}
which is predicted to be $\alpha = -1$ in the heavy $c$ quark limit. 
The subleading corrections to the heavy $c$ quark limit have been 
estimated and found to be small \cite{ML93}
\begin{equation}
\alpha < -0.95 ,
\end{equation}
and recent measurements yield 
\begin{eqnarray} 
\alpha &=& -0.91 \pm 0.49 \quad \mbox{ARGUS \cite{ARGUSalpha}} \\
\alpha &=& -0.89^{+0.17+0.09}_{-0.11-0.05}  \quad \mbox{CLEO\cite{CLEOlamc}}
\end{eqnarray}
and are in satisfactory agreement with the theoretical predictions. 

Recently the CLEO collaboration also measured the ratio of the form factors
$F_1$ and $F_2$, averaged over phase space. Heavy quark symmetries do not 
fix this form factor ratio, at least not for a heavy to light decay, while 
for a heavy to heavy decay the form factor $F_2$ vanishes in the heavy mass 
limit for the final state quark. CLEO measures \cite{CLEOratio}
\begin{equation}
\left\langle \frac{F_2}{F_2} \right\rangle_{\mbox{phase space}} = 
     -0.25 \pm 0.14 \pm 0.08
\end{equation}
which is in good agreement with model estimates \cite{JKlambda}.

\section{The $1/m_Q$ Expansion in Inclusive Decays}
For inclusive decays a $1/m_Q$ expansion is obtained for the 
rates by an approach similar to the one known from deep inelastic
scattering \cite{Manifesto}-\cite{FLS93}. 
The first step consists of an operator product expansion (OPE)
which yields an infinite sum of operators with increasing dimension. 
The dimensions of the operators are compensated by inverse powers 
of a large scale, which is in general of the order of the heavy mass 
scale. The decay probability is then given as forward matrix elements
of these operators between the state of the decaying heavy hadron; 
these matrix elements still have a mass dependence, which then may 
be extracted in terms of a $1/m_Q$ expansion using HQET as for exclusive 
decays. 

The method described below also allows us to deal with inclusive 
non-leptonic processes and hence in principle opens the possibility for 
a calculation of lifetimes and branching fractions in the framework of 
the $1/m_Q$ expansion. This is remarkable, since non-leptonic processes
are usually very hard to deal with, in particular the $1/m_Q$ expansion 
has not (yet ?) brought any success in the field of exclusive non-leptonic 
decays.

Applying the OPE to the energy spectra of the charged lepton in 
inclusive semileptonic decays of heavy mesons, the relevant expansion
parameter is not $1/m_Q$, but rather $1/(m_Q - 2E_\ell  )$; the 
denominator is thus the energy release of 
the decay. In almost all phase space the energy release is of 
the order of the heavy mass; it is only in the endpoint region that it
becomes small and hence the expansion breaks down. This problem 
may be fixed by a resummation of terms in the operator product 
expansion, which strongly resembles the summation corresponding to 
leading twist in deep inelastic scattering. Analogously to the 
parton-distribution function, a universal function appears, which 
determines all inclusive heavy-to-light decays.

\subsection{Operator Product Expansion for Inclusive Decays}
The effective Hamiltonian for a decay of a heavy 
(down-type) quark is in general linear in the decaying heavy 
flavoured quark  
\begin{equation} \label{heff}
{\cal H}_{eff} = \bar{Q} R
\end{equation}
where the operator $R$ describes the decay products. 
In the following we shall consider semileptonic decays, for which  
\begin{equation} \label{rsl}
R_{sl} = \frac{G_F}{\sqrt{2}} V_{Qq} \,\, \gamma_\mu (1-\gamma_5) q 
\,\, (\bar{\nu}_\ell \gamma^\mu (1-\gamma_5) \ell)  , 
\end{equation}
where $q$ is an up-type quark ($c$ or $u$, since we shall consider $b$ 
decays). Similarly, for non-leptonic decays the Cabbibo allowed contribution 
corresponds to  
\begin{eqnarray}
R_{nl} &=& \frac{G_F}{2\sqrt{2}} V_{Qq} V^*_{q'q''} \left[ (C_+ (m_b) + C_- (m_b))  
         \gamma_\mu (1-\gamma_5) q  \right. (\bar{q}' \gamma^\mu (1-\gamma_5) q'')
       \\ 
       && \qquad \qquad \qquad + (C_+ (m_b) - C_- (m_b)) 
         \gamma_\mu (1-\gamma_5) q''  
       \left. (\bar{q}' \gamma^\mu (1-\gamma_5) q )\right], 
\nonumber 
\end{eqnarray}
where $q'$ ($q''$) is a  down-type (up-type) quark and $V_{Qq}$ the corresponding 
CKM matrix element. The coefficients $C_\pm (m_b)$
are the QCD corrections obtained from the renormalization group running between 
$M_W$ and $m_b$; in leading logarithmic approximation these coefficients are \cite{C12} 
\begin{equation} \label{cpm}
C_{\pm} (m_b) =[\frac{\alpha _S(M_W^2)}{\alpha _S(m_b^2)}]^{\gamma_{\pm}},
\mbox{ with } \gamma _+=\frac{6}{33-2N_f}=-\frac{1}{2}\gamma _-
\end{equation}
where $\alpha_s (\mu)$ is the onle-loop expression (\ref{alpha}) for the running coupling 
coupling constant of QCD.

Finally, for  radiative rare decays we have
\begin{equation}
R_{rare} = \frac{G_F}{\sqrt{2}} V_{tb} V^*_{ts} C_7 (m_b) \frac{e}{16 \pi^2} 
            m_b \sigma_{\mu \nu} (1+\gamma_5) s F^{\mu \nu} 
\end{equation} 
where $C_7 (m_b) $ is again a coefficient obtained from running between $M_W$ 
and $m_b$. Its value is $C_7 (m_b) \sim 0.3$, the corresponding analytical 
expression may be found in \cite{C7}.  

The inclusive decay rate for a heavy hadron $H$ containing the quark $Q$ 
may be related to a forward matrix element by  
\begin{eqnarray} 
&& \Gamma  \propto  \sum_X (2 \pi)^4 \delta^4 (P_B - P_X )
| \langle X | {\cal H}_{eff} | H(v) \rangle |^2 
= \label{inclusive}
\int d^4 x \,   \langle H(v) |{\cal H}_{eff} (x) 
            {\cal H}_{eff}^\dagger (0) | H(v) \rangle 
\\ \nonumber
&&=  2 \mbox{ Im}
\int d^4 x \,   \langle H(v) |T \{ {\cal H}_{eff} (x) 
            {\cal H}_{eff}^\dagger (0) \} | H(v) \rangle .
\end{eqnarray}
where $| X \rangle $ is the final state, which is summed over to 
obtain the inclusive rate.

The matrix element appearing in (\ref{inclusive}) contains a 
large scale, namely the mass of the heavy quark. The first step 
towards a $1/m_Q$ expansion is to make this large scale explicit. 
This may be  done by a phase redefinition. This leads to 
\begin{equation}
\Gamma  \propto
2 \mbox{ Im} 
\int d^4 x  \, e^{-im_Q vx} \,  
\langle H(v) |T \{ \widetilde{{\cal H}}_{eff} (x) 
   \widetilde{{\cal H}}_{eff} ^\dagger (0) \} | H(v) \rangle 
\end{equation}
where 
\begin{equation} 
\widetilde{{\cal H}}_{eff}
= \bar{Q}_v R \qquad Q_v = e^{-im_Q vx} Q 
\end{equation}
This relation exhibits the similarity 
between the cross-section calculation in deep inelastic scattering 
and the present approach to total rates. In deep inelastic scattering 
there appears a large scale, which is the momentum transfer to the 
leptons, while here the mass of the heavy quark appears as a large scale. 

The next step is to perform an operator product expansion of the product 
of the two Hamiltonians. After the phase redefinition 
the remaining matrix element does not involve large momenta of the order 
of the heavy quark mass any more and hence a short-distance expansion  
becomes useful, if the mass $m_Q$ is large compared to the scale 
$\bar\Lambda$ determining the matrix element.
The next step is thus to perform an operator-product 
expansion, which has the general form  
$$
\int d^4 x  \, e^{im_Q vx} \, 
\langle H(v) |T \{ \widetilde{{\cal H}}_{eff} (x) 
\widetilde{{\cal H}}_{eff}^\dagger (0) \}| H(v) \rangle
= \sum_{n=0}^\infty  \left(\frac{1}{2m_Q}\right)^n 
     \hat{C}_{n+3} (\mu) \langle H(v) |{\cal O}_{n+3}| H(v) \rangle_\mu ,
$$
where ${\cal O}_n$ are operators of dimension $n$, with their 
matrix elements renormalized at scale 
$\mu$, and $\hat{C}_n$ are the corresponding Wilson coefficients. 
These coefficients encode the short distance physics related to the 
heavy quark mass scale and may be calculated in perturbation theory. 
All long distance contributions connected to the hadronic scale 
$\bar\Lambda$ are contained in the matrix elements of the operators 
${\cal O}_{n+3}$.      

Still the matrix elements of ${\cal O}_{n+3}$ are not independent of the 
heavy quark mass scale, but this mass dependence may be expanded in powers
of $1/m_Q$ by means of heavy quark effective theory. This is achived
by expanding the heavy quark fields appearing in the operators ${\cal O}_n$
using (\ref{HarvField}) (or, equivalently, (\ref{FWfield}))
as well as the states by including the corrections to the Lagrangian  
given in (\ref{HarvLag}) (or (\ref{FWlag})) 
as time-ordered products. In this way the 
mass dependence of the total decay rate may be accessed completely within 
an expansion in $1/m_Q$.  

The lowest-order term of the operator product expansion are the  
dimension-3 operators. Due to Lorentz invariance and parity there 
are only two combinations which may appear, namely   
$\bar  Q_v \fmslash{v} Q_v$ or $\bar Q_v Q_v$. Note that the $Q_v$
operators differ from the full QCD operators only by a phase redefinition, 
and hence $\bar  Q_v \fmslash{v} Q_v = \bar  Q \fmslash{v} Q$ and 
$\bar Q_v Q_v = \bar Q Q$.  The first combination 
is proportional to the $Q$-number current $\bar  Q \gamma_\mu Q$, 
which is normalized even in full QCD,
while the second differs from the first one only by terms of order
$1/m_Q^2$ 
\begin{equation}
\bar Q_v Q_v = v_\mu \bar Q_v \gamma_\mu Q_v 
+ \frac{1}{2 m_Q^2}
\bar{h}_v  \left[ (iD)^2 - (ivD)^2 +
\frac{i}{2} \sigma_{\mu \nu} G^{\mu \nu} \right] h_v  
+ {\cal O} (1/m_Q^3) .
\end{equation}
where $G_{\mu \nu}$ is the gluon field strength.

Thus the matrix elements of the dimension-3 contribution 
is known to be normalized; in the standard normalization of the 
states this implies 
\begin{equation}
\langle H(v) |{\cal O}_3| H(v) \rangle = 
\langle H(v) |\bar Q_v \fmslash{v} H_v | B(v) \rangle = 2 m_H 
\end{equation}
where $m_H$ is the mass of the heavy hadron.
To lowest order in the heavy mass expansion we may furthermore 
replace $m_B = m_Q$ and hence we may evaluate the leading term in 
the $1/m_q$ expansion without any hadronic uncertainty. Generically   
the dimension-3 contribution yields the free quark decay rate. 
This has been previously used as a model for inclusive decays, but
now it turns out to be the first term in a systematic $1/m_Q$ expansion 
of total rates. 

A dimension-four operators contains an additional covariant derivative, 
and thus one has matrix elements of the type 
\begin{equation}
\langle H(v) |{\cal O}_4| H(v) \rangle \propto 
\langle H(v) |\bar Q_v \Gamma D_\mu Q_v | H(v) \rangle = {\cal A}_\Gamma v_\mu  
\end{equation}
Since the equations of motion apply for this tree level matrix element, 
one finds that the constant ${\cal A}_\Gamma$ has to vanish, and thus 
there are no dimension-four contributions. This statement is completely 
equaivalent to Lukes theorem \cite{Lu91}, since we are considering 
a forward matrix element, i.e. a matrix element at zero recoil \cite{zerec}.

The first non-trivial non-perturbative contribution comes from
dimension-5 operators and are of order $1/m_Q^2$.  
For mesonic decays there are only the two parameters $\lambda_1$
and $\lambda_2$ given in (\ref{lam1}) and (\ref{lam2}), 
which correspond to matrix elements of the subleading terms of 
the Lagrangian. They parametrize the non-perturbative input in the order 
$1/m_Q^2$. For $\Lambda_Q$-type baryons the parameter $\lambda_2$ 
vanishes due to heavy quark spin symmetry, while the kinetic energy parameter 
$\lambda_1$ is non-zero as well. In the framework of the $1/m_Q$ 
expansion this leads to a difference in lifetimes between mesons and
baryons.

\subsection{Calculation of Total Decay Rates}
In this subsection we shall collect the results for the total rates 
including the first non-trivial non-perturbative correction. 

Inserting $R_{sl}$ as given in (\ref{rsl}) one obtains for the total 
inclusive semileptonic decay rate $B \to X_c \ell \nu$
\begin{equation} \label{bcsl}
\Gamma(B \to X_c \ell \nu)   
 = \frac{G_F^2 m_b^5}{192\pi^3} |V_{cb}|^2 
      \left[
      \left(1+\frac{\lambda_1}{2m_c^2}\right) f_1 \left(\frac{m_c}{m_b}\right)
      -\frac{9\lambda_2}{2m_c^2} f_2 \left(\frac{m_c}{m_b}\right)
      \right] , 
\end{equation}
where the two $f_j$ are phase-space functions 
\begin{eqnarray} \label{f1}
f_1(x) &=& 1-8x^2+8x^6-x^8-24x^4\log x ,
\\  \nonumber
f_2(x) &=& 1-\frac{8}{3}x^2-8x^4+8x^6+\frac{5}{3}x^8+8x^4\log x  .
\end{eqnarray}
The result for $B \to X_u \ell \nu_\ell$ is obtained from (\ref{bcsl}) 
as the limit $m_c \to 0$ and the replacement $V_{cb} \to V_{ub}$   
\begin{equation} \label{busl}
\Gamma(B \to X_u \ell \nu) = \frac{G_F^2 m_b^5}{192\pi^3} |V_{ub}|^2
 \left[
      1+\frac{\lambda_1 - 9 \lambda_2}{2m_b^2} 
      \right]   .
\end{equation}
As it has been discussed above, the leading non-perturbative 
corrections in (\ref{bcsl}) and (\ref{busl}) are parametrized 
by $\lambda_1$ and $\lambda_2$. Estimates for these parameters 
have been discussed in section 2; in order to 
estimate the total effect of the non-perturbative effects we shall 
insert a range of values $-0.3 > \lambda_1 >  -0.6$ GeV${}^2$; from this 
we obtain   
\begin{equation}
\frac{\lambda_1 - 9 \lambda_2}{2m_b^2} \sim -(3 \cdots 4) \%
\end{equation}
This means that the non-perturbative contributions are small, 
in particular compared to the perturbative ones, which have been 
calculated some time ago \cite{QCDradcorr,ACCMM}. For the decay 
$B \to X_u \ell \bar\nu_\ell$ the lowest order QCD corrections 
are given by 
\begin{equation}
\Gamma (B \to X_u \ell \bar{\nu}_\ell) =  \frac{G_F^2 m_b^5}{192\pi^3} 
       |V_{ub}|^2 \left[ 1 + \frac{2 \alpha}{3 \pi}
                    \left(\frac{25}{4}-\pi^2\right) \right]
= 0.85 |V_{ub}|^2 \Gamma_b,  
\end{equation}
and thus the typical size of QCD radiative corrections is of the order of 
ten to twenty percent. 

Similarly, one obtains the result for non-leptonic decays as 
\begin{equation}
\Gamma (B\to X_c) = 3 \frac{G_F^2 m_b^5}{192\pi^3}
\left\{ A_1 f_1 \left(\frac{m_c}{m_b} \right) 
\left[ 1 + \frac{1}{2 m_b^2} (\lambda_1 - 9 \lambda_2) \right] 
        - 48 A_2 f_3 \left(\frac{m_c}{m_b}\right)
          \frac{1}{2m_b^2} \lambda_2 \right\} 
\end{equation}
where the coefficients $A_i$ are given by combination of the Wilson coefficents 
$C_\pm (m_b)$ (\ref{cpm})
\begin{equation}
A_1 = \frac{1}{3} [C_-^2 (m_b) + 2 C_+^2 (m_b)] , \qquad
A_2 = \frac{1}{6} [C_+^2 (m_b) - C_-^2 (m_b)] ,
\end{equation}
and $ f_3 (x) = (1-x^2)^3 $ is another phase space function.  
Again the non-perturbative corrections turn out to be small, in the 
region of a few percent compared to the leading term, and the perturbative
corrections turn out to be much larger than this. 

Finally, for the rare decay $B \to X_s \gamma$ one may as well calculate 
the non-perturbative contribution in terms of $\lambda_1$ and $\lambda_2$.
One obtains  
\begin{equation}  
\Gamma (B\to X_s\gamma)
     =  \frac{\alpha G_F^2}{16\pi^4}
     m_b^5 |V_{ts} V_{td}^*|^2 |C_7(m_b)|^2  
    \left[1+\frac{1}{2m_b^2}
    \left(\lambda_1-9\lambda_2 \right)\right],  
\end{equation}
and the relative size of the non-perturbative corrections is the 
same as in the $B \to X_u \ell \bar\nu_\ell$ decays. 

Typically the non-pertubative corrections are much smaller than the radiative
corrections. The only exception is the endpoint region of lepton energy spectra
which receives both large perturbative as well as non-perturbative corrections. 
However, this is only a small region in phase space and the corections to the 
total rates remain moderate.    

\subsection{Lifetimes of Heavy Hadrons}
The subject of heavy hadron lifetimes is strongly related to non-leptonic 
processes, which have been considered already some time ago 
\cite{rueckl}-\cite{CabbMaia}; however the application of the $1/m_Q$ 
expansion has turned many assumptions into quantitative arguments. 
As outlined in the last section the $1/m_Q$ expansion allows to calculate
total rates, even for non-leptonic processes, and hence a QCD-based calculation 
of lifetimes becomes possible. A recent review of this subject is given in 
\cite{bigilife}.

Studying the formulae obtained in the $1/m_Q$ expansion up to order $1/m_Q^2$ 
one finds that lifetime differences between $B$ mesons do not occur up to this 
level of the expansion; in other words, any difference between the $B^\pm$ and 
the $B^0$ or the $B_s$ lifetimes are induced by effects of the order $1/m_b^3$. 
These effects are due to dimension six operators of the four-quark form 
\begin{equation} \label{4quark}
{\cal O}_6 = (\bar b \Gamma q) (\bar q \Gamma b) .
\end{equation}

\begin{figure}
\begin{center}
--------------- FIG 1 -------------------
\end{center}
   \caption{\label{fig4}
Diagrams for non-leptonic decays of $B$ mesons. The upper diagram
corresponds to the leading term in the $1/m_Q$ expansion, the middle
one to weak anihilation, and the lower one to Pauli interference. 
Diagrams taken from \protect\cite{bigilife}.}
\end{figure}

In fig.\ref{fig4} the relevant diagramms are shown. The two lower ones 
yield dimension six operators of the from (\ref{4quark}); comparing to the 
terminology used in phenomenological models the middle diagram is called
weak annihilation (WA) and the lower one Pauli interference (PI). 

The four quark operators appearing at order $1/m_Q^3$ are usually estimated
applying the vacuum insertion assumption, althought this procedure
has been criticized recently \cite{NeuSach}. It amounts to 
replace 
$$ 
\langle B | (\bar b \Gamma q) (\bar q \Gamma b) | B \rangle \to   
\langle B | (\bar b \Gamma q) | 0 \rangle \langle 0 |(\bar q \Gamma b) | B \rangle 
$$
and hence the parameter entering the estimates of the lifetime differences
is the decay constant $f_B$. 

The WA piece has been considered in \cite{PG,BUWA} and has been found 
to be small, of the order of one percent. The dominant contribution comes 
from the PI diagram. For the lifetime of the $B^-$ one obtains \cite{bigilife}
\begin{equation}
\Gamma (B^-)=\Gamma _{1/m_Q^2} (B) + \Delta \Gamma _{PI}(B^-)
\end{equation}
where the PI contribution is
\begin{equation}
\Delta \Gamma _{PI}(B^-) =  \frac{G_F^2m_b^5}{192\pi ^3}|V_{cb}|^2
\cdot
24\pi ^2 \frac{f_B^2}{M_B^2}[C_+^2(m_b) -C_-^2(m_b) 
      +\frac{1}{N_C}(C_+^2(m_b) +C_-^2(m_b))]
\end{equation}
where $C_\pm$ have been given in (\ref{cpm}). 
QCD radiative corrections change the sign of the PI contribution; the constructive
interference at the scale $M_W$ is turned into a destructive one at the scale $m_b$
prolonging the lifetime of the $B^-$ relative to the $B_0$. Using HQET
the running of the coefficients below $m_b$ has also been calculated;
it has been found that the destructive interference is amplified by the running 
below $m_b$, since
\begin{eqnarray}
\Delta \Gamma _{PI}(B^-) = \frac{G_F^2m_b^5}{192\pi ^3}|V_{cb}|^2 \cdot 
24\pi ^2 \frac{f_B^2}{M_B^2}
\kappa ^{-4}&& [(C_+^2(m_b)-C_-^2(m_b))\kappa ^{9/2}+\frac{C_+^2(m_b)+C_-^2(m_b)}{3}
\nonumber \\ 
&& - \frac{1}{9}(\kappa ^{9/2}-1)(C_+^2(m_b)-C_-^2(m_b))], 
\end{eqnarray}
where 
\begin{equation}
\kappa = [\frac{\alpha _S(\mu ^2_{had})}{\alpha_S(m_b^2)}]^{1/b}, \mbox{ with }
b=11-\frac{2}{3}n_F
\end{equation}
and $\mu_{had}$ is some small hadronic scale, which may be defined e.g.\ by 
$\alpha_s (\mu_{had}) = 1$. Putting in numbers the result is \cite{bigilife}
\begin{equation} \label{lrat}
\frac{\tau (B^-)}{\tau (B_d)} =
1+0.05\cdot \frac{f_B^2}{(200\; \mbox{MeV} )^2}
\end{equation}
Hence the $1/m_Q$ expansion (in combination with vacuum insertion) predicts a
slightly longer lifetime for the charged $B$ meson compared to the neutral one.  
However, this is based on the vacuum insertion assumption which has been 
reconsidered recently in \cite{NeuSach}; it is claimed that possible 
non-factorizable contributions may be enhanced by large prefactors thereby 
invalidating (\ref{lrat}).

However, the estimate (\ref{lrat}) is compatible with data; the 
latest compilation yields \cite{Sciolla}
\begin{equation}
\frac{\tau (B^+)}{\tau (B^0)}=1.019 \pm 0.048
\end{equation}

Applying the $1/m_Q$ expansion for the lifetimes also to the $D$ system  
one finds large corrections. From the experimental side it is knwon that 
the lifetime differences are very large and that the naive parton model 
expectation fails by a large margin. This may be taken as an indication 
that the $c$ quark mass is indeed not large enough to justify a $1/m_Q$ 
expansion, at least for the non-leptonic processes.      

The lifetimes of the two neutral $B$ mesons are equal up to terms of order
$1/m_b^3$, hence one expects 
\begin{equation}
\bar \tau (B_d) = \bar \tau (B_s) 
\end{equation}
to a good accuracy. Here $\bar \tau$ denotes the average lifetime 
of the two mass eigenstates of the $B^0-\bar B^0$ system. While the 
lifetimes in the $B_d$ system are practically the same such that the 
difference may be neglected, it has been pointed out that this is not 
necessarily the case in the $B_s$ system; the lifetime difference 
has been estimated in \cite{bigibsosc} to be     
\begin{equation}
\frac{\Delta \Gamma (B_s)}{\bar \Gamma (B_s)}
= \frac{\Gamma (B_{s,short})-\Gamma (B_{s,long})}
{\bar \Gamma (B_s)} =  0.18 \cdot
\frac{f_{B_s}^2}{(200\; \mbox{MeV} )^2}
\end{equation}
Hence this difference may be as large twenty percent and thus it 
cannot be neglected any more e.g. in an analysis of $B_s$-$\bar{B}_s$
mixing. 

Finally we shall consider the $b$ baryon lifetimes. Here we expect 
the differences between the meson and the baryon lifetimes to be of 
order $1/m^2$, since the matrix elements of the kinetic energy operator
as well as of the chromomagnetic moment operator are different between 
baryons and mesons; in particular, due to spin symmetry 
the matrix element of the chromomagnetic moment operator vanishes for 
$\Lambda$-type baryons, since the light degrees of freedom are in a spin-0
state. Probably only the $\Lambda$-type baryons will be stable against 
electromagnetic and strong decays, and hence their lifetime will 
be determined by the weak decay of the heavy quark.   

Theoretical estimates of the $\Lambda_b$ lifetime have been attempted 
in \cite{bigilife, NeuSach} and yield the expectation that 
\begin{equation}
\frac{\tau (\Lambda _b)}{\tau (B_d)}
\stackrel{>}{\sim} 0.9 .
\end{equation}
This has to be confronted with the recent data \cite{Sciolla}
\begin{equation}
\frac{\tau (\Lambda _b)}{\tau (B_d)} = 0.80 \pm 0.05 .
\end{equation}
Thus the lifetime is slightly below the expectation, although this is 
not yet a significant deviation from the prediction. In particular, 
older data indicated that this lifetime ratio could have been as low
as 0.7; such a low value would clearly indicate a theoretical problem
in the $1/m_Q$ expansion of the lifetimes, but the new data are in better 
agreement with the theoretical expectations.

\subsection{Lepton Energy Spectra}
The method of the operator-product expansion may also  be used to 
obtain the non-perturbative corrections to the charged lepton energy 
spectrum \cite{CCGincsl}. In this case the operator product expansion is 
applied not to the full effective Hamiltonian, but rather only to the 
hadronic currents. The rate is written as a product of the hadronic and
leptonic tensor
\begin{equation}
d \Gamma = \frac{G_F^2}{4 m_B} | V_{Qq} |^2 W_{\mu \nu}  
\Lambda^{\mu \nu} d(PS) , 
\end{equation}
where $d(PS)$ is the phase-space differential.  
The short-distance expansion is then performed for 
the two currents appearing in 
the hadronic tensor. Redefining the phase of the heavy-quark fields as in 
(\ref{phaseredef}) one finds that the momentum transfer variable 
relevant for the short-distance expansion is $m_Q v - q$, where 
$q$ is the momentum transfer to the leptons. 
  
The structure of the expansion for the spectrum is identical to the one 
of the total rate. The contribution of the dimension-3 operators 
yields the free-quark decay spectrum, there are no contributions from 
dimension-4 operators, and the $1/m_b^2$ corrections are parametrized 
in terms of $\lambda_1$ and $\lambda_2$. Calculating the spectrum for 
$B \to X_c \ell \nu$ yields \cite{Bigiincsl}-\cite{tmincsl}
\begin{eqnarray}
\frac{d \Gamma}{dy} &=& \frac{G_F^2\,|\,V_{cb}|^2\,m_b^5}{192\pi^3}
\Theta(1-y-\rho) y^2 \left[\left\{ 3 (1-\rho) (1-R^2) - 2 y (1-R^3) \right\}
\right. \\ 
&& + \frac{\lambda_1}{[ m_b(1-y)]^2} (3 R^2-4 R^3) 
  - \frac{\lambda_1}{m_b^2(1-y)} (R^2-2 R^3) 
\nonumber \\
&& - \frac{3\lambda_2}{m_b^2(1-y)} (2 R+3 R^2-5 R^3) 
   + \frac{\lambda_1}{3 m_b^2} [ 5y - 2(3-\rho) R^2 + 4 R^3 ] 
\nonumber\\
&& \left. + \frac{\lambda_2}{m_b^2} [ (6+5y) - 12 R
    - (9-5\rho) R^2 + 10 R^3 ] \right] 
    + {\cal O}\left[ (\Lambda/[m_b(1-y)])^3 \right]
\nonumber  
\end{eqnarray}
where we have defined 
\begin{equation}
\rho = \left(\frac{m_c}{m_b}\right)^2 \quad  R = \frac{\rho}{1-y}
\end{equation}
and 
\begin{equation}
y = 2 E_\ell / m_b 
\end{equation}
is the rescaled energy of the charged lepton. 

This expression is somewhat complicated, but it simplifies 
for the decay $B \to X_u \ell \nu$ since then the mass of the quark 
in the final state may be neglected. One finds
\begin{eqnarray} \label{buspec}
\frac{d\Gamma}{dy} &=&  \frac{G_F^2\,|\,V_{ub}|^2\,m_b^5}{192\pi^3}
        \left[ \left( 2y^2 (3-2y)
        + \frac{10y^2}{3} \frac{\lambda_1}{m_b^2}
        + 2y(6+5y) \frac{\lambda_2}{m_b^2} \right) \Theta(1-y) \right. 
\nonumber \\
&&  \vphantom{\frac{G_F^2\,|\,V_{qb}|^2\,m_b^5}{192\pi^3}}
\left. - \frac{\lambda_1 + 33 \lambda_2}{3m_b^2} \delta (1-y)
  -  \frac{\lambda_1}{3m_b^2} \delta ' (1-y) \right]  
\end{eqnarray}

\begin{figure}
\begin{center}
------------- FIG 2 -----------------
\end{center}
    \caption{The electron spectrum for free quark
$b\rightarrow c$ decay (dashed line), free quark $b\rightarrow u$ decay
(grey line), and  $B\rightarrow X_c e \bar\nu_e$ decay including
$1/m_b^2$ corrections (solid line) with $\lambda_1 = - 0.5$ GeV${}^2$
and $\lambda_2 = 0.12$ GeV${}^2$. The figure is from
\protect\cite{mwincsl}.} 
\label{fig2}
\end{figure}

Figure~\ref{fig2} shows the distributions for inclusive semileptonic 
decays of $B$ mesons. The spectrum close to the endpoint, where the 
lepton energy becomes maximal, exhibits a sharp spike as $y \to y_{max}$. 
In this region we have we have 
\begin{equation}  
\frac{d\Gamma}{dy}  \propto \Theta(1-y-\rho)
\left[ 2 + \frac{\lambda_1}{(m_Q (1-y))^2} 
\left(\frac{\rho}{1-\rho} \right)^2 
\left\{ 3 - 4 \left(\frac{\rho}{1-\rho} \right) \right\}  \right] , 
\end{equation} 
which behaves like 
$\delta$-functions and its derivatives as $\rho \to 0$, which can be 
seen in (\ref{buspec}). This behaviour indicates a breakdown of the operator 
product expansion close to the endpoint, since for the spectra the 
expansion parameter is not $1/m_Q$, but rather $1/(m_Q - qv)$, which 
becomes $1/(m_Q [1-y])$ after the integration over the neutrino momentum. 
In order to obtain a description of the endpoint region, one has to perform
some resummation of the operator product expansion.

\subsection{Resummation in the Endpoint Region}
Very close to the endpoint of the inclusive semileptonic decay spectra
only a few resonances contribute. In this resonance region one may 
not expect to have a good description of the spectrum using
an approach based on parton-hadron duality; here a sum over a few
resonances will be appropriate. 

In the variable $y$ the size of this resonance region is however of the 
order of $(\bar\Lambda / m_Q)^2$ and thus small. In a larger region of 
the order $\bar\Lambda / m_Q $, which we shall call the endpoint region, 
many resonances contribute and one may hope to describe the spectrum 
in this region using parton-hadron duality. 

It has been argued in \cite{NeubertShape} that the $\delta$-function-like
singularities appearing in (\ref{buspec}) may be reinterpreted as 
the expansion of a non-perturbative function describing the spectrum in the 
endpoint region. Keeping only the singular
terms of (\ref{buspec}) we write
\begin{equation} \label{shape}
\frac{1}{\Gamma_b} \frac{d\Gamma}{dy}  = 2y^2 (3-2y) S(y) ,
\end{equation}
where 
\begin{equation} \label{momexp}
S(y) = \Theta (1-y) + \sum_{n=0}^{\infty} a_n \delta^{(n)} (1-y)
\end{equation}
is a non-perturbative function given in terms of the moments $a_n$ of the 
spectrum, taken over the endpoint region. These moments themselves have an 
expansion in $1/m_Q$ such that $a_n \sim 1/m_Q^{n+1}$, 
and we shall consider only the leading term in 
the expansion of the moments, corresponding to the most singular 
contribution to the endpoint region. 

Comparing (\ref{buspec}) with 
(\ref{shape}) and (\ref{momexp}) one obtains that
\begin{eqnarray} 
a_0 &=& \int dy (S(y) - \Theta (1-y)) = {\cal O} (1/m_Q^2)  \\ \label{a1}
a_1 &=& \int y  (S(y) - \Theta (1-y)) = -\frac{\lambda_1}{3 m_Q^2}
\end{eqnarray} 
where the integral extends over the endpoint region. 

The non-perturbative function implements a resummation of the 
most singular terms contributing to the endpoint and, in the language
of deep inelastic scattering, corresponds to the 
leading twist contribution. This 
resummation has been studied in QCD \cite{ManNeu,BiMotion} and the 
function $S(y)$ may be related to the distribution of the light cone 
component of the heavy quark residual momentum 
inside the heavy meson. The latter is a fundamental 
function for inclusive heavy-to-light transitions, which has been 
defined in \cite{BiMotion}
\begin{equation}\label{fdef}
   f(k_+) = \frac{1}{2M_H} 
   \langle H(v)|\,\bar h_v\,\delta(k_+-i D_+)\,h_v\,
   |H(v)\rangle ,       
\end{equation} 
where $k_+ = k_0 + k_3$ is the positive light cone component
of the residual momentum $k$. The relation between the two functions
$S$ and $f$ is given by
\begin{equation} 
S(y) = \frac{1}{m_Q} \int\limits_{-m_Q (1-y)}^{\bar\Lambda} dk_+ f(k_+)
\end{equation}
from which we infer that the $n^{th}$ moment of the endpoint region
is given in terms of the matrix element $\langle H(v)| 
\bar h_v (i D_+)^n h_v |H(v)\rangle $. 

The function $f$ is a universal distribution function, which 
appears in all heavy-to-light inclusive decays; another example 
is the decay $B \to X_s \gamma$ \cite{NeubertBsg,BiMotion}, 
where this function determines 
the photon-energy spectrum in a region of order $1/m_Q$ around
the $K^*$ peak. 

In principle $f$ has to be determined by other methods than the $1/m_Q$
expansion, e.g.\ from lattice calculations or from a model, or it has 
to be determined from experiment by measuring the photon spectrum 
in $B \to X_s \gamma$ or the lepton spectrum in $B \to X_u \ell \bar{\nu}$.
In the context of the model ACCMM model \cite{ACCMM} $f$ has been 
calculated in \cite{BigiACCM}.  

Some of the properties of $f$ are known. Its support is 
$-\infty < k_+ < \bar\Lambda$, it is normalized to unity, and its 
first moment vanishes. Its second moment is given by $a_1$, and its 
third moment has been estimated \cite{BiMotion,zerec}. A one-parameter 
model for $f$ has been suggested in \cite{ManNeu}, which incorporates 
the known features of $f$
\begin{equation} \label{ftoy}
   f(k_+) = {32\over\pi^2\bar\Lambda}\,(1-x)^2 
   \exp\bigg\{ - {4\over\pi}\,(1-x)^2 \bigg\}\,
   \Theta(1-x)  ,
\end{equation}
where $x = k_+ / \bar\Lambda$, and   
the choice $\bar\Lambda = 570$ MeV yields reasonable values 
for the moments. In fig.~\ref{fig3} we show the spectrum for 
$B \to X_u \ell \nu_\ell$ using the ansatz (\ref{ftoy}). 

\begin{figure}
\begin{center}
------------- FIG 3 ------------------
\end{center}
   \caption{Charged-lepton spectrum in $B\to X_u \ell  \bar\nu$ 
decays. The solid line is (\protect{\ref{shape}}) with the ansatz 
(\protect{\ref{ftoy}}), the dashed
line shows the prediction of the free-quark decay model. The figure 
is from \protect\cite{ManNeu}.}
\label{fig3}
\end{figure}

Including the non-perturbative effects yields a reasonably behaved
spectrum in the endpoint region and the $\delta$-function-like 
singularities have disappeared. Furthermore, the spectrum now extends
beyond the parton model endpoint;   
it is shifted from $E_\ell^{max} = m_Q / 2$ to the physical 
endpoint $E_\ell^{max} = M_H / 2$, since $f$ is non-vanishing for 
positive values of $k_+ < \bar\Lambda = M_H - m_Q$.

\section{Conclusions}
The development of the field of heavy quark physics has been indeed
remarkable over the last few years, experimentally as well as theoretically.
From the experimental side,  the progress in the technology 
of detectors (e.g.\ silicon vertex detectors) opened the possibility to 
study $b$ physics even at machines which originally were not designed 
for this kind of research. In this way also the high energy colliders 
(in particular LEP and TEVATRON) could contribute substantially in this 
area, since they allow to measure states (such as the $B_s$ and the $b$
flavoured baryons) which lie above the threshold of the 
$\Upsilon (4s)$-$B$-factories. 

From the theoretical side the heavy quark limit and HQET brought an 
important success, since it provides a model independent and QCD based 
framework for the description of processes involving heavy quarks. 
The effective theory approach has originally been formulated for 
exclusive decays but in the past few years a heavy mass expansion has 
been set up also for inclusive transitions. 

As far as exclusive heavy to heavy decays are concerned, the additional 
symmetries of the heavy mass limit restrict the number of 
non-perturbative functions in a model independent way; furthermore, 
heavy quark symmetries fix the absolute normalization of some of the 
transition amplitudes at the point of maximum momentum transfer. 
Phenomenologically this has improved our knowledge on the CKM matrix
element $V_{cb}$ dramatically; with the value $|V_{cb}| = (39.5 \pm 2.0)
\times 10^{-3}$ the relative precision of this CKM matrix element is now
about 5\% and thus at a level of the precision with which the Cabbibo 
angle is known. 

In heavy to light 
decays heavy quark symmetries do not work as efficiently; in this case
only the relative normalization of $B$ decays versus the corresponding 
$D$ decays may be obtained. From the experimental side there are first 
measurements of $B \to \pi \ell \nu$ and $B \to \rho \ell \nu$ from the 
CLEO collaboration and 
an extraction of the CKM matrix element $V_{ub}$ from these processes
is still to some extent model dependent. The latest value for this 
CKM matrix element 
is $V_{ub} = (3.3 \pm 0.2 ^{+0.3}_{-0.4} \pm 0.7) \times 10^{-3}$
where the last error is due to the model dependence.

HQET does not yet have much to say about exclusive non-leptonic decays; even 
for the decays $B \to D^{(*)} D^{(*)}_s$, which involves three heavy quarks, 
heavy quark symmetries are not sufficient to yield useful relations between
the decay rates \cite{MRRnonlep}. Of course, with additional assumptions 
such as factorization one can go ahead and relate the 
non-leptonic decays to the semileptonic ones; however, this is a very strong
assumption and it is not clear in what sense factorization is an 
approximation. On the other side, the data on the non-leptonic $B$ decays 
support factorization, and first attempts to understand this from QCD and 
HQET have been untertaken \cite{DGnonlep}; however, the problem of the 
exclusive non-leptonic decays still needs clarification and hopefully the 
heavy mass expansion will also be useful here.

The $1/m_Q$ expansion obtained from the OPE and HQET offers the unique 
possibility to calculate the transition rates for inclusive decays in a QCD 
based and model independent framework. The leading term of this expansion 
is always the free quark decay, and the first non-trivial corrections are 
in general given by the mean kinetic energy of the heavy quark inside the 
heavy hadron $\lambda_1$ and the matrix element $\lambda_2$ of the 
chromomagnetic moment operator.

The method also allows us to calculate differential distributions, such as
the charged lepton energy spectrum in inclusive semileptonic decays of heavy 
hadrons. For this case, the expansion parameter is the inverse of the energy 
release $m_b - 2 E_\ell$, where $E_\ell$ is the lepton energy. Close to the 
endpoint, the energy  release is small and thus the expansion in its
inverse powers becomes useless. In this kinematic region one may partially 
resume the $1/m_Q$ expansion, obtaining a result closely analogous to the 
leading twist term in deep inelastic scattering. Particularly in the endpoint 
region a non-perturbative function is needed which corresponds to the parton 
distributions parametrizing the deep inelastic scattering.
 
The leading term of the $1/m_Q$ expansion is the free quark decay, and the 
result for the semileptonic branching fraction in this approximation has been 
well known for some time. The first non-perturbative corrections turn out
to be quite small and hence the main corrections are the perturbative 
QCD corrections, where in a recent calculation also the effects of finite 
charm quark mass have been taken into account \cite{Ball,Ballnew}. The 
radiative corrections lower the semileptonic branching fraction somewhat 
compared to the parton model. 

There has been some discussion on the issue of the semileptonic branching 
fraction triggered by the fact that the data used to be as low as 
$Br(B \to X \ell \nu) \sim 10.5\%$ with a relative error of about ten percent. 
Such a low value for the semileptonic branching fraction in combination with 
tha charm counting in $B$ decays would indicate some theoretical problem; 
however, the recent LEP data yield a value of 
$Br(B \to X \ell \nu) = (11.5 \pm 0.3)\%$ which is compatible with the 
theoretical expectations. 

The expansion in powers of the inverse quark mass has become the standard 
tool in heavy quark physics and with the forthcomming experiments one 
may expect a strong improvement in our knowledge of the CKM sector of 
the SM, in particular a test of CP violation as it is encoded in the 
CKM matrix of the SM.

\section*{Acknowledgment}
I thank the organizers of the Schladming School for the invitation and 
for providing such a beautiful environment for this conference.

\newpage
\section*{Figure Captions}
\begin{itemize}
\item[\bf Figure 1]{Diagrams for non-leptonic decays of $B$ mesons. 
                    The upper diagram
                    corresponds to the leading term in the $1/m_Q$ expansion, 
                    the middle one to weak anihilation, and the lower one 
                    to Pauli interference. 
                    Diagrams taken from \protect\cite{bigilife}.}
\item[\bf Figure 2]{The electron spectrum for free quark
                    $b\rightarrow c$ decay (dashed line), free quark 
                    $b\rightarrow u$ decay
                    (grey line), and  $B\rightarrow X_c e \bar\nu_e$ decay 
                    including $1/m_b^2$ corrections (solid line) with 
                    $\lambda_1 = - 0.5$ GeV${}^2$
                    and $\lambda_2 = 0.12$ GeV${}^2$. The figure is from
                    \protect\cite{mwincsl}.}
\item[\bf Figure 3]{Charged-lepton spectrum in $B\to X_u \ell  \bar\nu$ 
                    decays. The solid line is (\protect{\ref{shape}}) with 
                    the ansatz (\protect{\ref{ftoy}}), the dashed 
                    line shows the prediction of the free-quark decay model. 
                    The figure is from \protect\cite{ManNeu}.} 

\end{itemize}
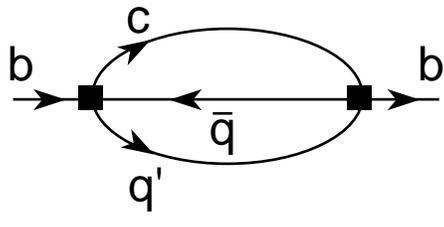
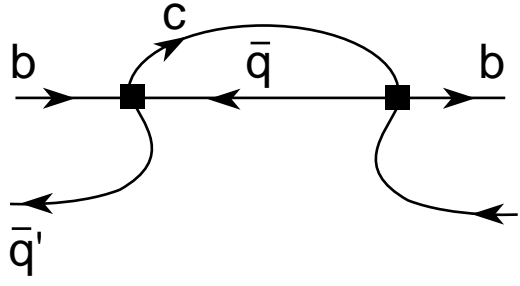
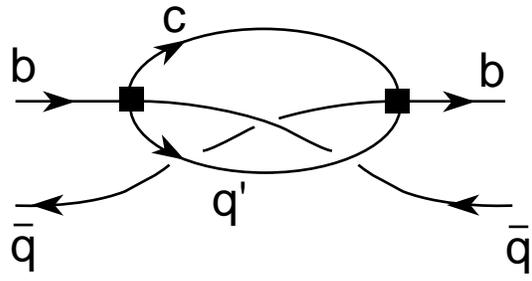
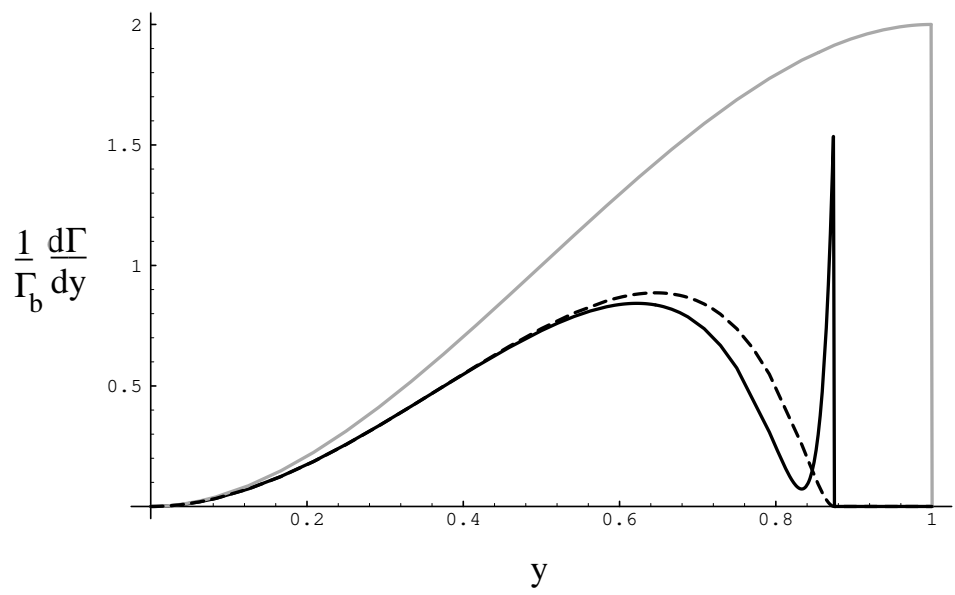
\end{document}